%% file: main.tex
\documentclass[aps,prd,twocolumn,showpacs,superscriptaddress]{revtex4-2}
\usepackage{graphicx}
\usepackage{tikz}
\usepackage{dcolumn}
\usepackage{bm}
\usepackage{hyperref}
\usepackage{multirow}

\begin{document}

\preprint{APS/123-QED}

\title{Cosmogenic Neutron Production in Water at SNO+}

\input{authorlist_2025-11-04}

\begin{abstract}
Accurate measurement of the cosmogenic muon-induced neutron yield is crucial for constraining a significant background in a wide range of low-energy physics searches. Although previous underground experiments have measured this yield across various cosmogenic muon energies, SNO+ is uniquely positioned due to its exposure to one of the highest average cosmogenic muon energies at $364\,\textup{GeV}$. Using ultra-pure water, we have determined a neutron yield of $Y_{n}=(3.38^{+0.23}_{-0.30})\times10^{-4}\,\textup{cm}^{2}\textup{g}^{-1}\mu^{-1}$ at SNO+. Comparison with simulations demonstrates clear agreement with the \textsc{FLUKA} neutron production model, highlighting discrepancies with the widely used \textsc{GEANT4} model. Furthermore, this measurement reveals a lower cosmogenic neutron yield than that observed by the SNO experiment, which used heavy water under identical muon flux conditions. This result provides new evidence that nuclear structure and target material composition significantly influence neutron production by cosmogenic muons, offering fresh insight with important implications for the design and background modelling of future underground experiments.
\end{abstract}

\maketitle

\section{\label{sec:Intro} Introduction}
Cosmogenic muons are produced in the showers initiated by cosmic rays interacting with the Earth's atmosphere. High-energy cosmogenic muons will penetrate the Earth's crust, reaching underground laboratories. These muons can produce neutrons and radioactive isotopes, which constitute a significant background to physics searches in underground experiments, such as neutrino, neutrinoless double beta decay and dark matter searches. Measuring the neutron yield of cosmogenic muons is important for understanding this background.

Cosmogenic neutron production arises from a variety of processes. The muons themselves can produce neutrons through direct interactions with nuclei, known as spallation. As muons propagate through matter, they lose energy via ionisation and radiative processes, producing secondary particles that induce electromagnetic and hadronic showers. Neutrons are generated both in photonuclear interactions within the electromagnetic showers, and through spallation processes involving secondary hadrons such as protons, pions, and other nuclear fragments \cite{wang_predicting_2001,nairat_neutron_2024}.

The number of neutrons associated with a muon depends on the muon's energy. Many underground experiments, with varying average muon energies, have measured the cosmogenic neutron yield. The relation between muon energy and neutron production was most recently demonstrated by Daya Bay \cite{an_cosmogenic_2018}, using liquid scintillator-based measurements \cite{hertenberger_muon-induced_1995,boehm_neutron_2000,blyth_measurement_2016,abe_production_2010, collaboration_measurement_1999,bellini_cosmogenic_2013}.

The average muon energy in an underground laboratory is dictated by the overburden. To date, the neutron yield measurement with the highest average muon energy was made by the Liquid Scintillator Detector (LSD) experiment \cite{aglietta_neutron_1989}. Measurements using muons of comparable average energies have since been made by the Sudbury Neutrino Observatory (SNO) experiment \cite{aharmim_cosmogenic_2019} and at the China Jinping Underground Laboratory \cite{zhao_measurement_2022}. The large overburden at SNO+ means the measurement in this paper will contribute to the high energy muon data.

The medium through which a muon propagates could also play a role in the number of neutrons produced. Many measurements of the neutron yield have been performed with liquid scintillator \cite{an_cosmogenic_2018,aglietta_neutron_1989,zhao_measurement_2022,abe_production_2010,boehm_neutron_2000,hertenberger_muon-induced_1995, blyth_measurement_2016,bellini_cosmogenic_2013,collaboration_measurement_1999}. SNO provided consistent results using both pure heavy water ($\textup{D}_{2}\textup{O}$) and NaCl-loaded heavy water. The first measurement using $\textup{H}_{2}\textup{O}$ came from Super-Kamiokande (SK), which used gadolinium-loaded water. The measurements by SK \cite{collaboration_measurement_2023}, in water, and KamLAND \cite{abe_production_2010}, in scintillator, were made using the same muon flux, providing a unique comparison of neutron production in different media. Their results were found to be consistent when normalised by medium density \cite{collaboration_measurement_2023}.

This paper will present the results on measuring the cosmogenic muon-induced neutron yield in ultra-pure water, using the SNO+ experiment. In water, cosmogenic neutrons will capture on hydrogen, causing the emission of a $2.2\,\textup{MeV}$ gamma ray. This gamma ray is the signal used for identifying neutrons. Detecting neutrons in a water Cherenkov detector is difficult because of the low energy signal of the neutron captures. Due to the very low backgrounds in the SNO+ detector, the trigger thresholds for detecting events can be set low enough to be able to detect neutrons in water. This measurement of the neutron yield is the first to be made using unloaded ultra-pure water. Further to this, SNO+ is the successor of the SNO experiment, receiving the same muon flux, providing the opportunity to make a direct comparison between the neutron yields measured using different media. 

\section{\label{sec:setup}Experimental Set-up}
SNO+ is a neutrino experiment that will search for neutrinoless double beta decay. Other physics goals of SNO+ include measurements of reactor, geo- and solar neutrinos. The detector infrastructure was inherited from the SNO experiment. While the SNO+ detector is currently filled with liquid scintillator, it was initially filled with ultra-pure water. During this period, known as the water phase, the electronics from SNO were re-commissioned and the background rates evaluated. The water phase was used as an opportunity to run the detector as a low-background water Cherenkov experiment. The water phase lasted from May 2017 to July 2019. Analyses from this phase include measurements of the $^{8}\textup{B}$ solar neutrino flux \cite{collaboration_measurement_2019,collaboration_measurement_2024}, nucleon decay searches \cite{Anderson_2019,collaboration_improved_2022}, and the first evidence of detecting reactor antineutrinos in a water Cherenkov detector \cite{collaboration_evidence_2023}.

\subsection{\label{subsec:SNO+_Detector}The SNO+ Detector}
\begin{figure}[ht]
  \centering
  \begin{tikzpicture}
    \node[anchor=south west,inner sep=0] (image) at (0,0) {\includegraphics[width=0.45\textwidth]{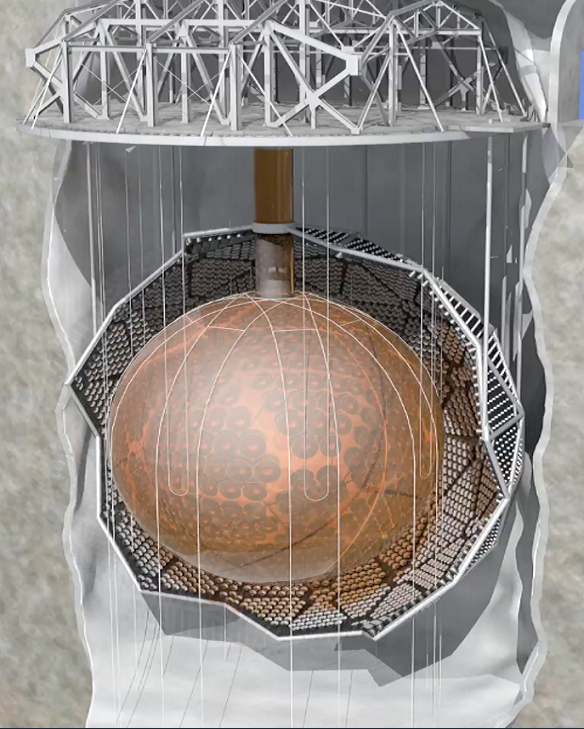}};
    \begin{scope}[x={(image.south east)}, y={(image.north west)}]
      \draw[-, ultra  thick, white] (0.8,0.1) -- (0.5,0.35);
      \draw[-, ultra  thick, white] (0.8,0.8) -- (0.75,0.6);
      \draw[-, ultra  thick, white] (0.2,0.1) -- (0.47,0.05);
      \draw[-, ultra  thick, white] (0.2,0.85) -- (0.48,0.75);
      \node[fill=white, draw=white, font=\small, text width=2cm] at (0.8,0.1) {AV filled with UPW};
      \node[fill=white, draw=white, font=\small, text width=2cm] at (0.8,0.8) {PMTs and OWLs mounted to the PSUP};
      \node[fill=white, draw=white, font=\small, text width=2cm] at (0.2,0.1) {Cavity filled with UPW};
      \node[fill=white, draw=white, font=\small, text width=1cm] at (0.2,0.85) {Neck};
    \end{scope}
  \end{tikzpicture}
  \caption{\label{fig:sno_plus_diagram}Diagram of the SNO+ detector with labelled components \cite{collaboration_sno_2021}.}
\end{figure}
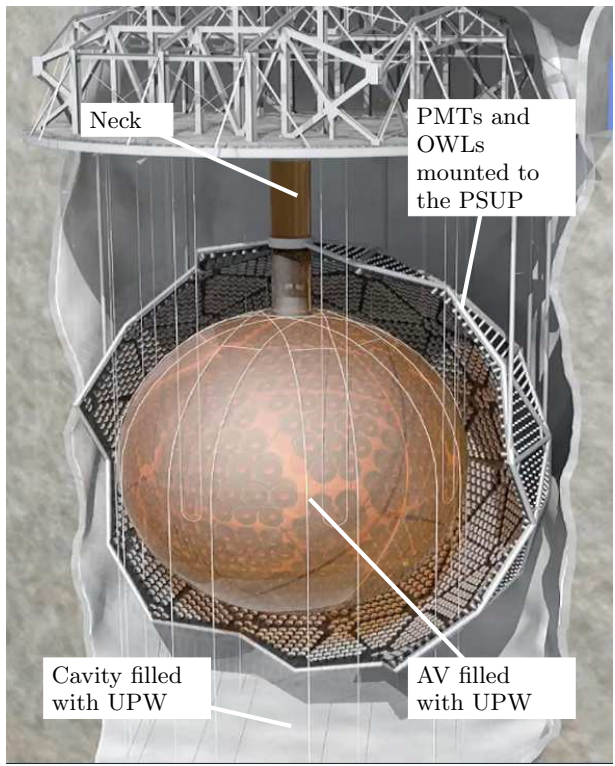

The SNO+ detector, shown in Figure~\ref{fig:sno_plus_diagram}, is located $2\,\textup{km}$ ($6010\,\textup{m.w.e.}$ in water equivalent meters) underground at SNOLAB \cite{lawson_snolab_2013,smith_snolab_2012}. An acrylic vessel (AV), of radius $6\,\textup{m}$ and thickness $5.5\,\textup{cm}$, contains the target medium that defines the phase. During the water phase, the AV was filled with $905\,\textup{tonnes}$ of ultra-pure water (UPW). A $7\,\textup{m}$ tall cylindrical neck is connected to the top of the AV, through which the AV can be filled and calibration sources can be deployed. At the top and facing down the neck are $4$ neck photomultiplier tubes (PMTs).

Surrounding the AV are 9362 inward-facing PMTs, all mounted on the stainless steel PMT support structure (PSUP). The PSUP has a radius of $8.9\,\textup{m}$. Also mounted on the PSUP are $91$ outward-looking PMTs (OWLs), which are used to identify cosmogenic muons using the Cherenkov light produced in the external water. The AV and the PSUP are positioned concentrically, and the space between them is filled with UPW. The entire setup is contained within a cavity filled with UPW that acts as a shield to radiation from the rock.

Part way through the water phase in September 2018, a cover gas system was installed to reduce the rate of $^{222}\textup{Rn}$ induced backgrounds in the AV. For this analysis, these decays contribute to the accidental coincidence backgrounds. This system uses nitrogen gas, which acts as a barrier to radon in the air above the liquid in the neck. A detailed description of the SNO+ detector and its cover gas systems can be found in \cite{collaboration_sno_2021}.

\subsection{\label{subsec:triggers}Event Triggers}
Each PMT has an adjustable channel threshold. When this threshold is exceeded, a hit is registered for that PMT, and a fixed-amplitude pulse of width $89\,\textup{ns}$ is produced. These pulses are summed across all PMTs. An event is recorded if the sum of the PMT pulses surpasses the event trigger threshold. For the water phase, the trigger threshold was set to $7$ pulses, approximately equivalent to a $1.4\,\textup{MeV}$ electron at the centre of the detector \cite{collaboration_measurement_2020}. The trigger system is modelled in SNO+ simulations, as described in Section~\ref{sec:Simulation}. %The resulting data is handled by the data acquisition systems, which prepare it for offline analysis.

% The trigger efficiency is the probability that an event will trigger the detector, given a number of PMT hits. Using simulation, it was found that the trigger efficiency is $100\,\%$ at and above $8$ PMT hits. Below $8$ PMT hits, the trigger efficiency decreases to zero. As a consequence, the corresponding total efficiency for triggering on a $2.2\,\textup{MeV}$ gamma ray, the signal for neutron capture, is approximately $50\,\%$ \cite{collaboration_measurement_2020}.

% The distribution of the summed PMT pulses can depend on the nature of the event. If the PMTs involved in an event all register hits at similar times, the sum of the pulses will form a sharper, more narrow peak than if the PMT hits were more spread over time. Close to the threshold, this event-dependence can affect whether an event triggers the detector or not. To minimise the impact of the trigger's dependency on the PMT hit distribution, the trigger efficiency is calculated as a function of `in-time hits', the number of PMT hits that arrive within an $89\,\textup{ns}$ time window, centred on the peak of the summed PMT pulses.
The trigger efficiency is the probability that an event will trigger the detector, given a number of PMT hits. The time distribution of the summed PMT pulses can depend on the nature of the event. If the PMTs involved in an event all register hits at similar times, the sum of the pulses will form a sharper, narrower peak than if the PMT hits were more spread over time. Close to the threshold, this event-dependence can affect whether an event triggers the detector or not. To minimise the impact of the trigger's dependency on the PMT hit time distribution, the trigger efficiency is calculated as a function of `in-time hits', the number of PMT hits that arrive within an $89\,\textup{ns}$ time window, centred on the peak of the summed PMT pulses. In \cite{collaboration_measurement_2020}, it was found using simulation that the trigger efficiency reaches $100\,\%$ at or above 8 PMT hits, with reduced efficiency below that threshold. For a 2.2 MeV gamma ray, the total efficiency was found to be approximately $50\,\%$ \cite{collaboration_measurement_2020}.

\subsection{\label{subsec:AmBe}Am-Be Calibration}
During the water phase, measurements of neutron captures in water were made using an americium-beryllium (Am-Be) calibration source. $^{241}\textup{Am}$ $\alpha$-decays with a half-life of $432\,\textup{years}$. These $\alpha$ particles can be absorbed by a $^{9}\textup{Be}$ nucleus to produce a $^{12}\textup{C}$ nucleus and a neutron. There is a $60\,\%$ probability that the $^{12}\textup{C}$ nucleus is in an excited state, which will immediately de-excite through the emission of a $4.4\,\textup{MeV}$ gamma ray. The neutrons typically capture on hydrogen nuclei, releasing a $2.2\,\textup{MeV}$ gamma ray.

The results of this calibration were presented in \cite{collaboration_measurement_2020}. The efficiency for detecting neutrons was found to be approximately $50\,\%$ near the centre of the detector. The neutron capture time was also measured to be $202.53^{+0.87}_{-0.76}\,\mu\textup{s}$. This neutron detection efficiency is relatively high for a water Cherenkov detector, allowing for the unique measurement of cosmogenic neutrons in UPW.

\section{\label{sec:Simulation}Simulation}
For this analysis, Monte Carlo (MC) simulations were performed for characterising the reconstruction of muons and neutrons, and measuring the efficiency for detecting and selecting neutrons. Simulations in SNO+ are handled by RAT, a software package used for offline analysis \cite{noauthor_rat_nodate}. All low level physics simulation and particle tracking is performed using the \textsc{GEANT4} (version 10.00.p04) software package \cite{agostinelli_geant4simulation_2003}, which has been incorporated into RAT, along with the detector geometry and trigger performance. Calibrations have been performed in order to tune and verify the full SNO+ simulation model.

Muons with a range of energies arrive at the detector from different directions. The angular distribution of cosmogenic muons at an underground site with a flat overburden is given by
\begin{equation}
    I(h)=\left(I_{1}e^{-h/\lambda_{1}}+I_{2}e^{-h/\lambda_{2}}\right)\sec{\theta},
    \label{eq:muon-depth-intensity-relation}
\end{equation}
where $\theta$ is the angle from zenith from which the muon arrives. The parameters used are $I_{1}=(8.60\pm0.53)\times10^{-6}\,\textup{cm}^{-2}\textup{s}^{-1}$, $I_{2}=(0.44\pm0.06)\times10^{-6}\,\textup{cm}^{-2}\textup{s}^{-1}$, $\lambda_{1}=0.45\pm0.01\,\textup{km.w.e.}$ and $\lambda_{2}=0.87\pm0.02\,\textup{km.w.e.}$, which have been determined using experimental data \cite{mei_muon-induced_2006}. The slant depth $h$ at SNO+ is given by $h=6.01\,\textup{km.w.e.}/\cos{\theta}$. The energy of these muons is described by
\begin{equation}
    \frac{dN}{dE}=Ae^{-bh(\gamma_{\mu}-1)}\left(E+\epsilon_{\mu}\left(1-e^{-bh}\right)\right)^{-\gamma_{\mu}},
    \label{eq:muon-energy-spec}
\end{equation}
where $A$ is the normalisation constant, $b=0.4\,/\textup{km.w.e.}$, $\gamma_{\mu}=3.77$ and $\epsilon_{\mu}=693\,\textup{GeV}$ \cite{mei_muon-induced_2006,groom_muon_2001}.

Muon simulations can either be run with the direction being sampled from an analytical expression, given by Equation~\ref{eq:muon-depth-intensity-relation} above, or with any given input direction. Whether the direction has been sampled or given as an input, the muon energy is then sampled from Equation~\ref{eq:muon-energy-spec}. The sampling of Equations~\ref{eq:muon-depth-intensity-relation}~and~\ref{eq:muon-energy-spec} gives an average muon energy of $(364\pm1)\,\textup{GeV}$ at SNO+.

\section{\label{sec:Muon_Reconstruction}Event Reconstruction}
The reconstruction of muons estimates the muon's direction and initial position on the PSUP. As muons travel through the detector, Cherenkov light is emitted, forming a cone around the particle's direction of travel. The high momentum of a cosmogenic muon means that the speed and direction of the muon remains approximately constant during its journey through the detector. The track of the muon is assumed to be a straight line between the muon's entry and exit points.

The inward-looking PMTs that are initially hit, as the muon enters the detector, provide information on the entry point. The muon's entry position is found by projecting the average position of the early hit PMTs onto the inner PSUP. As the muon exits the detector, the cone of Cherenkov light is focussed around the muon's exit point. The muon's exit position is estimated using the charge-weighted average of all hit PMTs. The direction of the muon is taken as the vector between these two points. More details on this procedure can be found in \cite{liggins_cosmic_2020}.

\begin{figure}
    \centering
    \includegraphics[width=\linewidth]{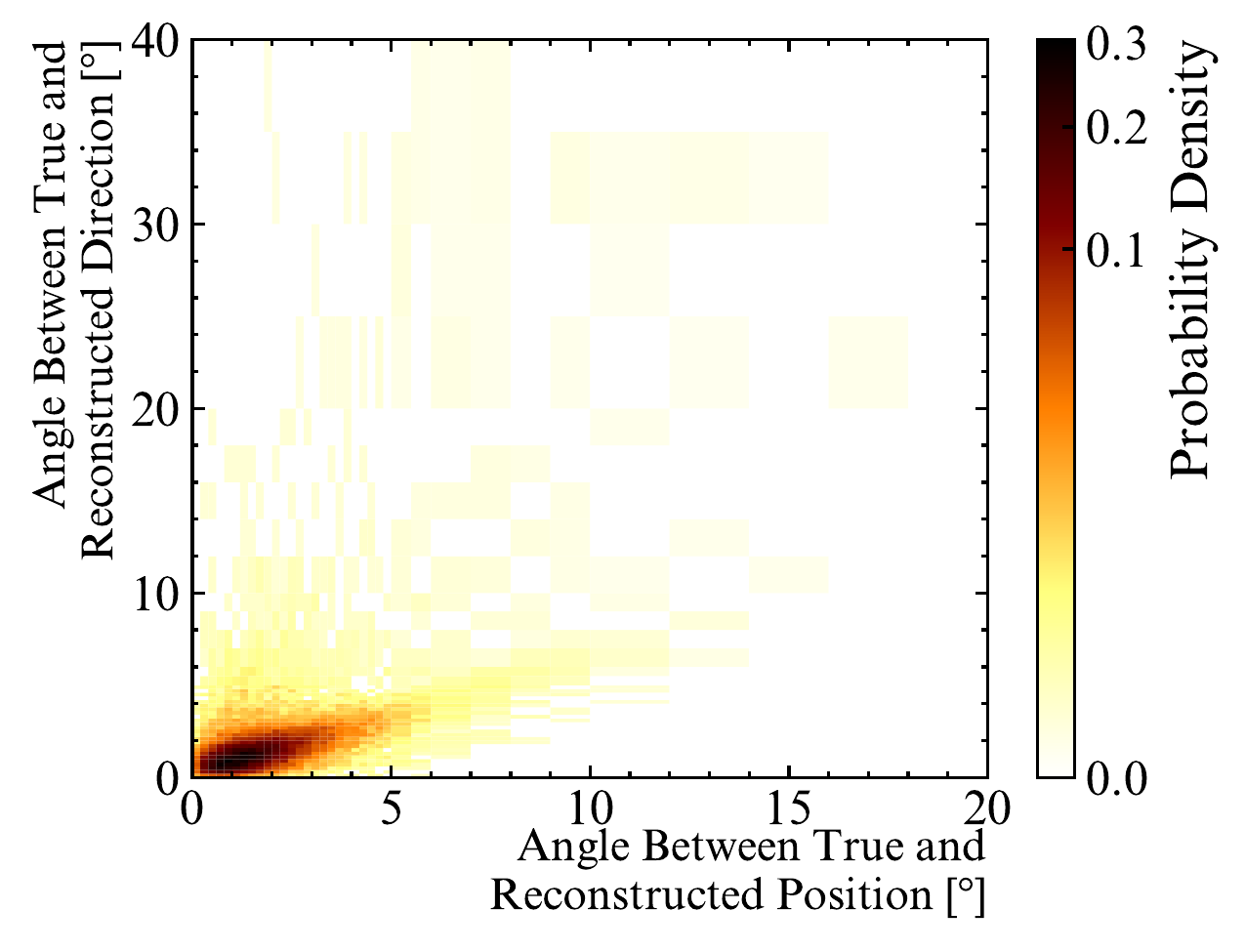}
    \caption{\label{fig:muonFitter}The angles between the true and reconstructed vectors for the directions and initial positions of simulated cosmogenic muons in SNO+.}
\end{figure}

Figure~\ref{fig:muonFitter} shows the reconstruction performance for cosmogenic muons. This distribution was created using simulated muons, where the truth is taken as the simulation itself, and the reconstructed information comes from applying the muon reconstruction to the simulation.

While the above describes the method for reconstructing muons, the reconstruction of the neutron capture signal follows a different method. The $2.2\,\textup{MeV}$ gamma ray from the neutron capture will Compton-scatter to produce secondary electrons, which produce Cherenkov radiation. These electrons have up to $1\,\textup{cm}$ track lengths in water, which is small compared to the $\sim20\,\textup{cm}$ position reconstruction resolution for low-energy events in the SNO+ water phase \cite{Anderson_2019}. The neutron capture signal is therefore treated as a point-like event. A likelihood approach is used to obtain the most probable event position, based on PMT hit times, utilising the spherical geometry of the detector. For an assumed event vertex, the distribution of PMT hit time residuals (the difference between the measured PMT hit times and the expected photon arrival times) is calculated. The likelihood is constructed by comparing the observed distribution of time residuals to the expected distribution, obtained from simulation, and is maximised with respect to the event position.

\section{\label{sec:Muon_Selection}Data Selection}
This analysis uses data taken during the water phase of SNO+, both before and after the installation of the cover gas system. The total run time across this period is $321\,\textup{days}$. Muon and neutron events are selected from the data using a suite of specifically designed selection criteria. Many of the selection criteria have been optimised using simulation with the goal of prioritising purity over efficiency, as presented in \cite{liggins_cosmic_2020}.

Due to the large overburden, only muons with high enough energies make it to the SNO+ detector. Muons that enter and exit the detector are known as through-going muons, and make up most of the population. Muons that stop in the detector are called stopping muons. The muon selection criteria, summarised in Table~\ref{tab:muon-selection}, have been formulated to select only a pure sample of through-going muons.

All cosmogenic muons will travel through the cavity water, producing Cherenkov light that can be detected by the OWLs. (a) All selected cosmogenic muons must have $5$ or more OWL hits. (b) Muons produce a large amount of Cherenkov light, so events with at least $1238$ calibrated PMT hits are selected. This cut was based on the spectrum measured by SNO for cosmogenic muons, after high level cuts had been applied \cite{collaboration_measurement_2009}.

Muons continually produce Cherenkov radiation along their direction of travel, leading to a large charge deposit in the PMTs closest to the muon's exit point on the PSUP. The amount of charge deposited decreases for PMTs further away from the exit point, creating a wide distribution of PMT charges. (c) Events are removed if the RMS of their PMT charge distribution is less than $4.5$~gain-scaled~units~(GSU), where the charge of each PMT has been divided by the gain applied to the PMT. As for the distribution of PMT hit times, this is expected to form a distinctive peak, where most PMTs were hit at the same time. (d) Events are discarded if the RMS of their PMT hit time distribution is greater than $38\,\textup{ns}$.

(e) Further conditions are applied to remove events that are unlikely to be cosmogenic muons. These conditions are referred to as data cleaning (DC) cuts, and are as follows. (i) Events caused by light leaks from the top of the detector are removed by cutting any events with at least $2$ neck PMT hits. (ii) Breakdowns in the electronics which cause a burst of events are flagged and removed. (iii) Muons that decay within the detector, producing a Michel electron, have a distinct signature that is used to remove them. These events are expected to have $\geq200$ PMT hits followed by an event with $\geq100$ PMT hits within $20\,\mu\textup{s}$ of the first event. (iv) If two events occur within $3\,\mu\textup{s}$ of each other, known as retriggers, these events are flagged as potentially being part of an electronics breakdown, and are removed.

\begin{figure}
    \centering
    \includegraphics[width=\linewidth]{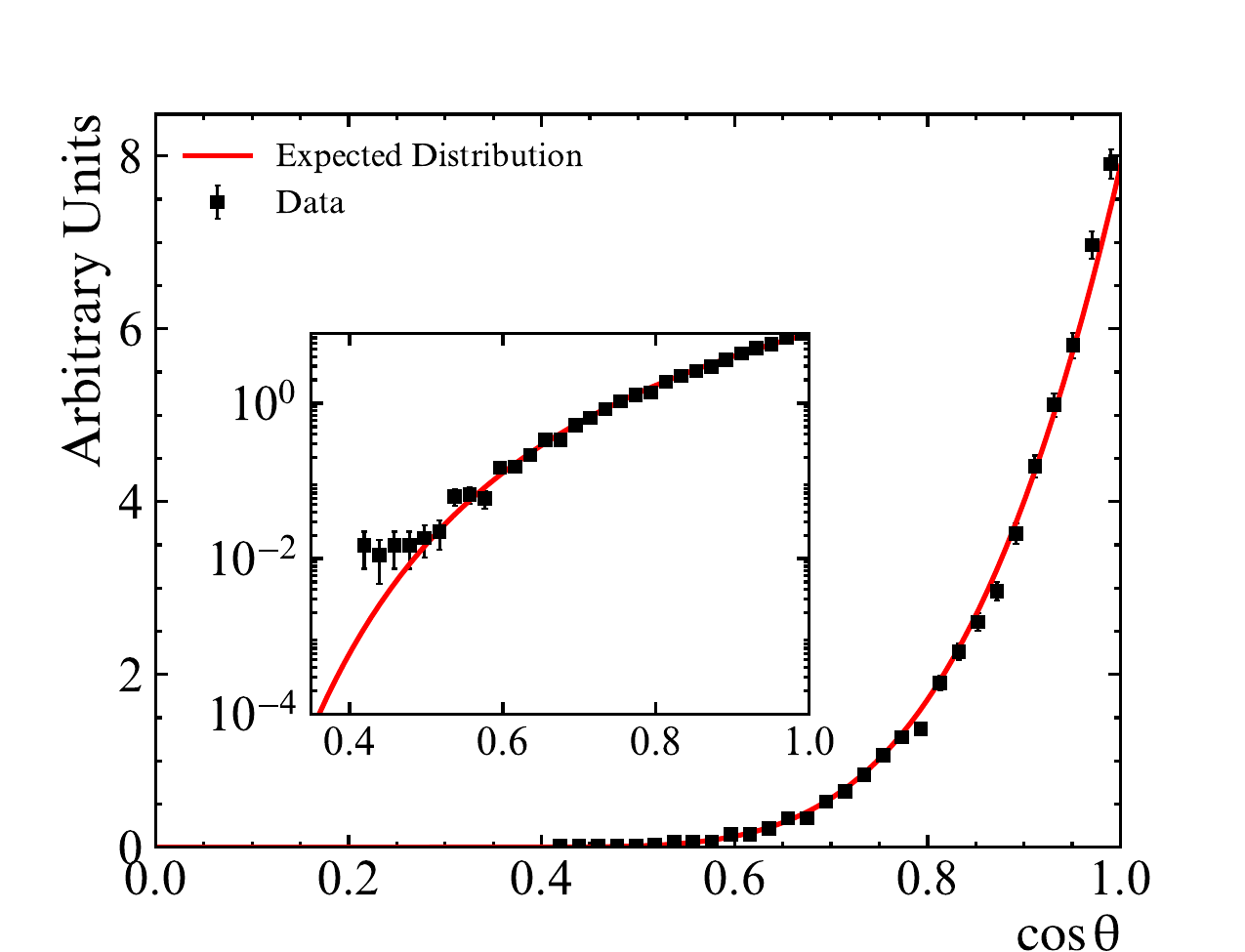}
    \caption{\label{fig:muon_dir}The zenith angle of the muon's direction $\theta$ for the muons selected from the data compared to the expected distribution, given by Equation~\ref{eq:muon-depth-intensity-relation}. Imbedded in the plot is a log-scale version.}
\end{figure}

As muon reconstruction requires an exit point, only through-going muons will be reconstructed. (f) The muon's reconstructed track length is required to be greater than $10\,\textup{m}$, where the track length is defined as the distance between the muon's entry and exit points on the PSUP. (g) As cosmogenic muons arrive at the detector from above, a condition that $\cos{\theta}\geq0.4$, where $\theta$ is the zenith angle of the muon's direction, is applied to ensure that the muons are down-going. Figure~\ref{fig:muon_dir} shows how the direction of the selected muons compares with the expected distribution from Equation~\ref{eq:muon-depth-intensity-relation}. There is small deviation at large incoming angles, which could be from atmospheric neutrino-induced muons, considered to be negligible.
\begin{table}[h]
\caption{\label{tab:muon-selection}Summary of the cosmogenic muon selection criteria.}
\begin{ruledtabular}
\begin{tabular}{ l c } 
        Parameter & Select if \\ \hline
        (a) OWL hits & $\geq5$ \\ \hline
        (b) Calibrated PMT hits & $\geq1238$ \\ \hline
        (c) Charge RMS & $\geq4.5\,\textup{GSU}$ \\ \hline
        (d) Time RMS & $\leq38\,\textup{ns}$ \\ \hline
        \multirow{4}{*}{(e) DC cuts} & (i) neck events \\
        & (ii) bursts \\ 
        & (iii) muon decay events \\
        & (iv) retriggers \\ \hline
        (f) Reconstructed track length & $\geq10\,\textup{m}$ \\ \hline
        (g) Direction & $\cos{\theta}\geq0.4$ \\
    \end{tabular}
\end{ruledtabular}
\end{table}

The neutron selection criteria are summarised in Table~\ref{tab:neutron-selection}. The neutron capture signal used for identifying neutrons creates a relatively low number of PMT hits. (a) Neutron follower events are selected if they satisfy $10\,\mu\textup{s}\leq\Delta T_{\mu}\leq776\,\mu\textup{s}$, where $\Delta T_{\mu}$ is the time since muon. (b) Of these events, those with in-time PMT hits less than $9$ and more than $25$ are removed. The lower timing and in-time hit conditions are to account for detector ringing, an effect that is often associated with a muon, when a large charge deposit results in a temporary effective shift in the trigger threshold. Further criteria are applied to the positions of the neutrons to reduce background contributions, which have been optimised through simulation. (c) Only neutron captures that have a perpendicular distance from their associated muon track of less than $4.6\,\textup{m}$ are selected, in order to reduce the contamination from random backgrounds within the detector. (d) Finally, the neutron capture position must have a radius of less than $8\,\textup{m}$, where the radius is the distance from the centre of the PSUP, which reduces the amount of external background contamination.
\begin{table}[h]
\caption{\label{tab:neutron-selection}Summary of the cosmogenic muon-induced neutron selection criteria.}
\begin{ruledtabular}
\begin{tabular}{ l c } 
Parameter & Select if \\ \hline
\multirow{2}{*}{(a) Time since muon} & $\geq10\,\mu\textup{s}$ \\ 
& $\leq776\,\mu\textup{s}$ \\  \hline
\multirow{2}{*}{(b) in-time PMT hits} & $\geq9$ \\ 
& $\leq25$ \\  \hline
(c) Distance from muon track & $<4.6\,\textup{m}$ \\  \hline
(d) Radius & $<8\,\textup{m}$
\end{tabular}
\end{ruledtabular}
\end{table}

\begin{figure}
    \centering
    \includegraphics[width=\linewidth]{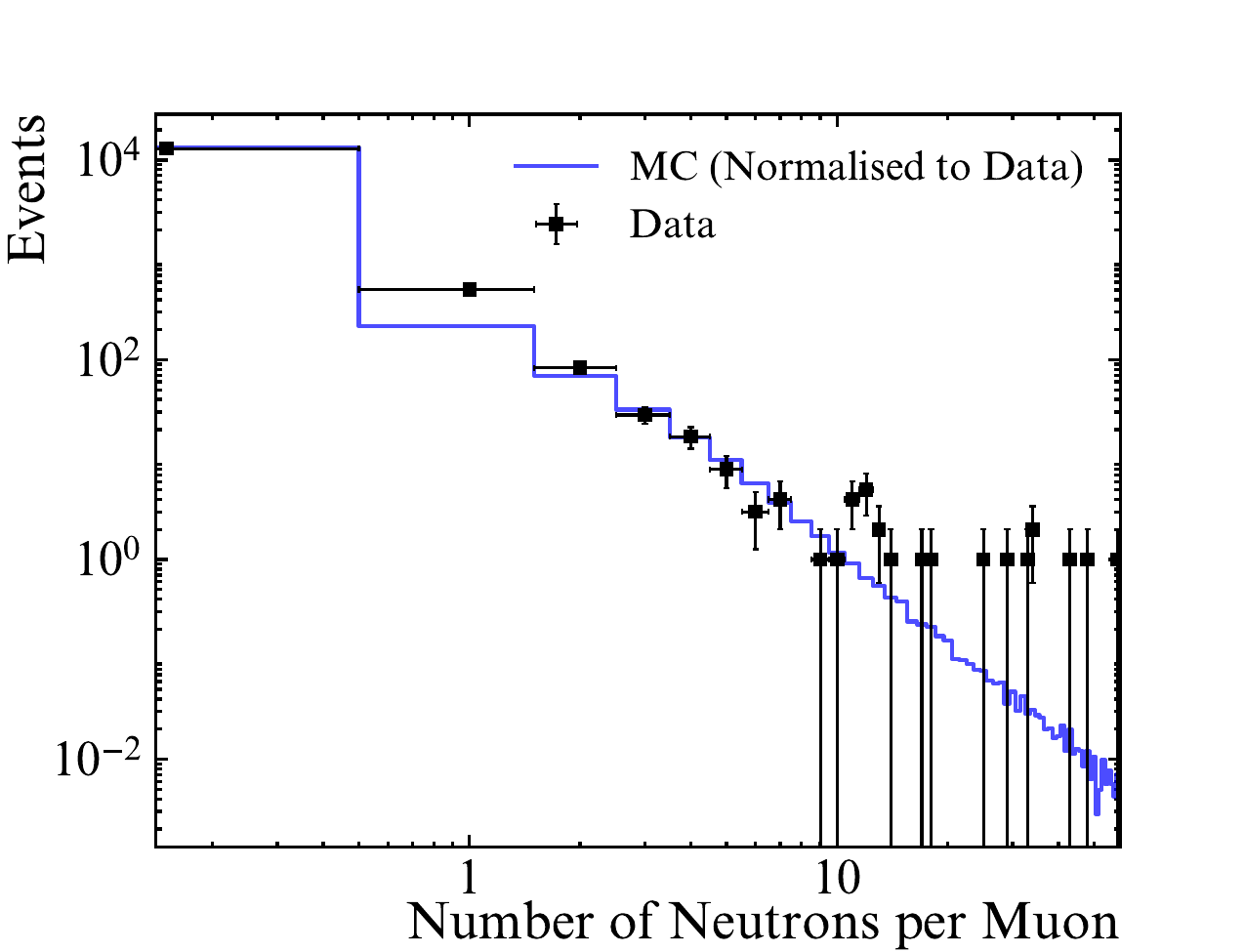}
    \caption{\label{fig:numNeutrons-dataMC}The number of neutrons selected per muon (black), compared with the number predicted in MC (blue).}
\end{figure}

\begin{figure}
    \centering
    \includegraphics[width=\linewidth]{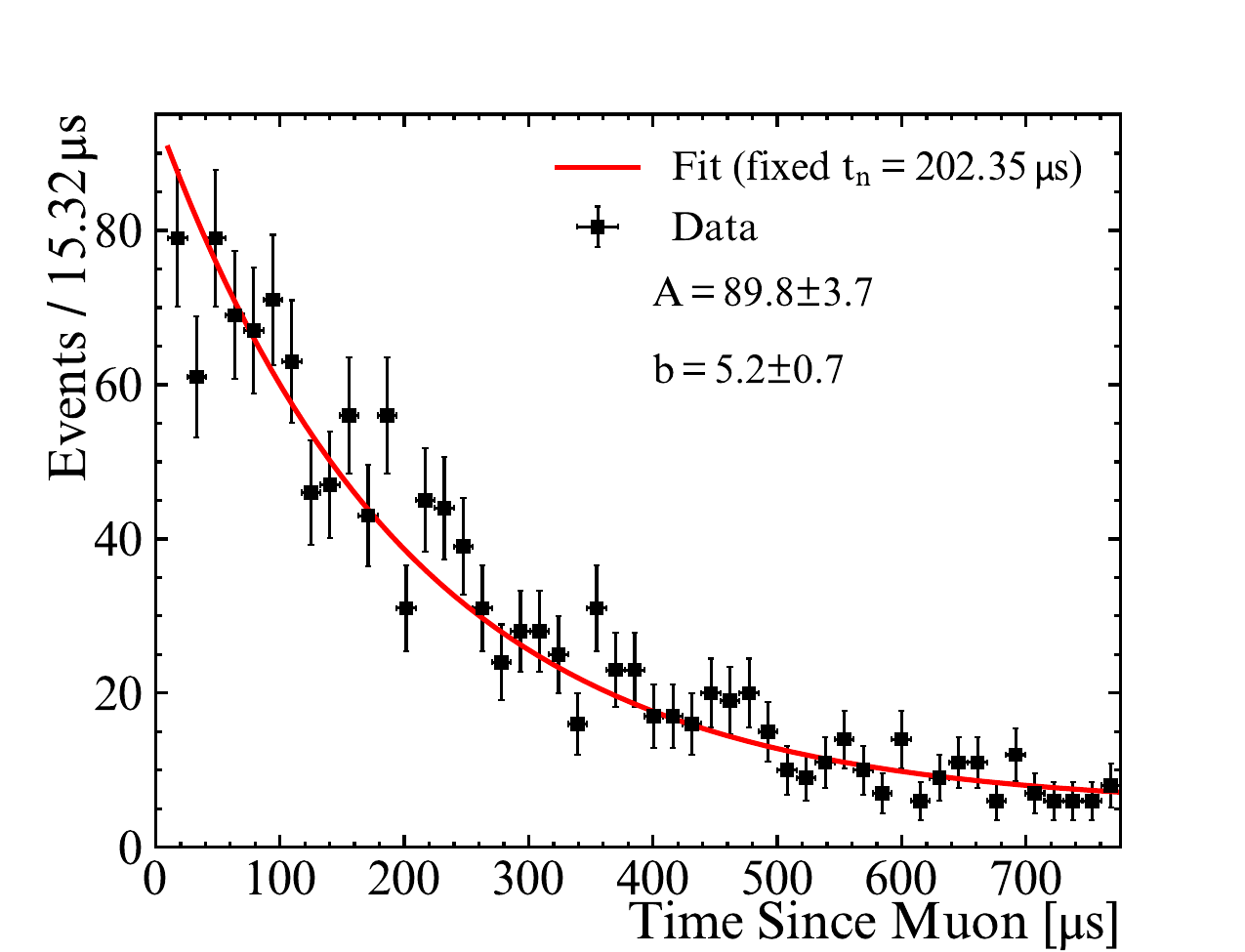}
    \caption{\label{fig:time_DATAvsMC}The time between the muon and neutron capture event (black), fitted with the neutron capture time measured during the Am-Be calibration (red) \cite{collaboration_measurement_2020}.}
\end{figure}

After applying the selection criteria outlined above, $13,690$ cosmogenic muons were selected with a total of $1,412$ neutron followers. Figure~\ref{fig:numNeutrons-dataMC} shows the number of selected neutron followers for each selected muon. The distribution of the times following the muon for the selected neutrons is shown in Figure~\ref{fig:time_DATAvsMC}. The data has been fitted with
\begin{equation}
    f(\Delta T_{\mu};b,A,t_{n}) = b + A\exp{\left(-\frac{\Delta T_{\mu}}{t_{n}}\right)},
    \label{eq:neutron-fit}
\end{equation}
where $b$ is the flat background component, $A$ is the amplitude, and $t_{n}$ is the neutron capture time, which has been set to the neutron capture time measured during the Am-Be water calibration \cite{collaboration_measurement_2020}.

\section{\label{sec:Method}Analysis Method}
\subsection{\label{sec:Neutron_Yield}Neutron Yield}
The neutron yield per muon ($Y_{n}$) is given by
\begin{equation}
    Y_{n} = \frac{N_{n}}{L\rho},
\label{eq:yield}
\end{equation}
where $N_{n}$ is the number of neutrons produced by the muon, $L$ is the track length of the muon, and $\rho$ is the density of the medium.

The number of neutrons produced per muon is related to the number of observed neutrons ($N_{\textup{obs}}$) via
\begin{equation}
    N_{\textup{obs}} = \varepsilon N_{n} + B_{C} + B_{R},
\label{eq:num_neutrons}
\end{equation}
where $\varepsilon$ is the neutron selection efficiency, $B_{C}$ is the muon correlated background count, and $B_{R}$ is the random detector background count. Both background contributions will be discussed in Section~\ref{sec:Backgrounds}, and the neutron selection efficiency will be discussed in Section~\ref{sec:Efficiency}.

\subsection{\label{sec:track_length}Muon Track Length}
The muon's track, along which neutron production is being measured, is defined by the two points where the muon track intersects the $8\,\textup{m}$-radius sphere defining the fiducial volume for this analysis. The length of the muon's track is equal to
\begin{equation}
    L = -2\vec{P}\cdot\vec{D},
    \label{eq:tarck_length}
\end{equation}
where $\vec{P}$ is the vector between the centre of the PSUP and the initial intersection point on the fiducial volume, and $\vec{D}$ is a unit vector in the muon's direction.

\begin{figure}
    \centering
    \includegraphics[width=\linewidth]{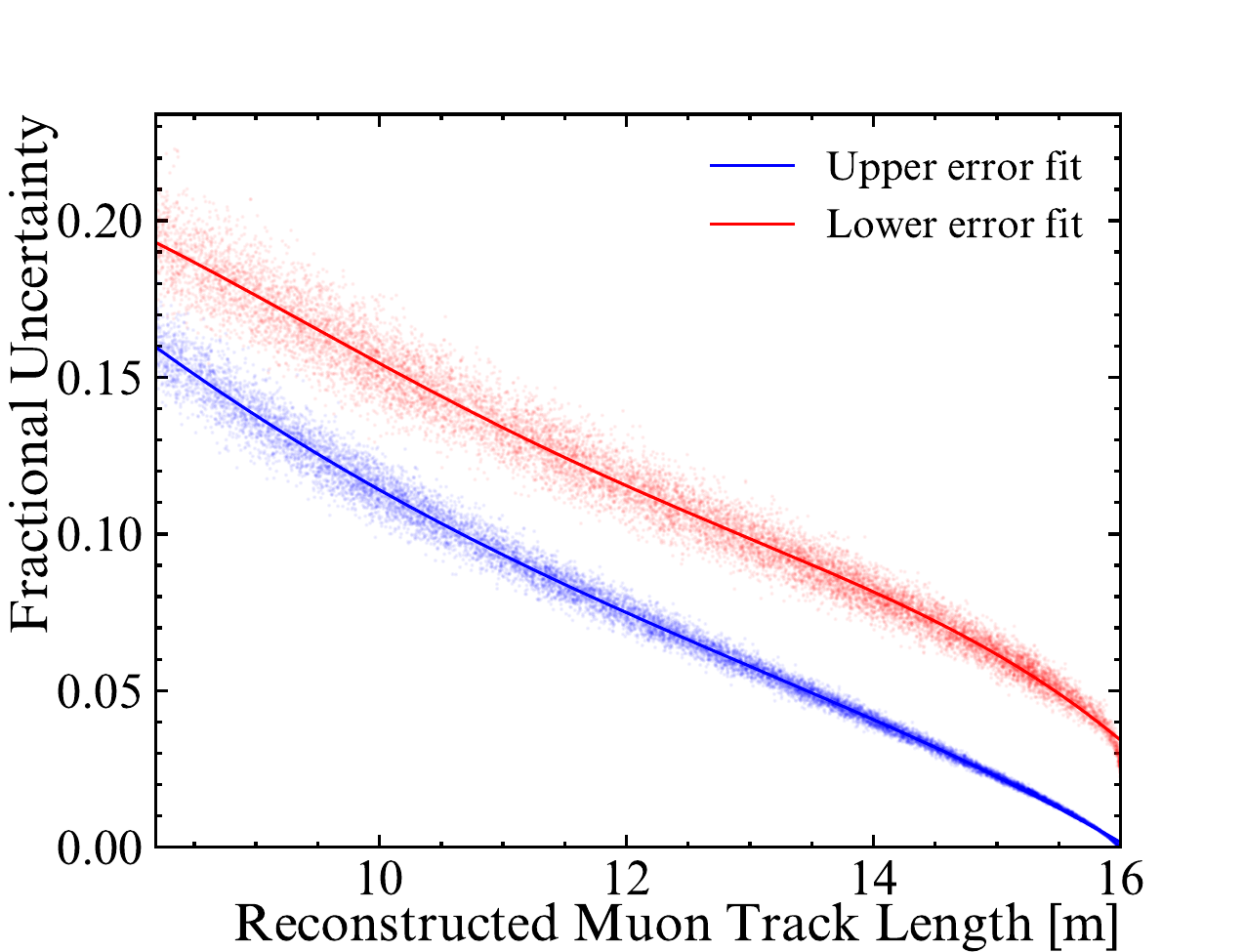}
    \caption{\label{fig:length_error}The fractional uncertainty in the reconstructed track length as a function of reconstructed track length. Due to the asymmetry of the uncertainty, it has been split into upper and lower errors, with each being fitted separately with a polynomial. }
\end{figure}

The uncertainty in the muon's initial position and direction both contribute to the track length uncertainty. Figure~\ref{fig:muonFitter} shows that these contributions are correlated. To include this correlation in the track length uncertainty, pairs of position and direction uncertainties are sampled from Figure~\ref{fig:muonFitter} and applied to each reconstructed muon in data, generating a set of perturbed track lengths. The mean deviation of these perturbed lengths from the original, divided by the original track length, defines the fractional uncertainty. These fractional uncertainties are plotted against reconstructed track length in Figure~\ref{fig:length_error} and fitted with a polynomial, which is then extrapolated to assign uncertainties to all muons in data.

\subsection{\label{sec:Backgrounds}Backgrounds}
One potential contribution to the background comes from other spallation products produced by the muon which mimic the neutron capture signal. This was evaluated using spallation isotope yields deduced from simulation \cite{li_first_2014}, which are compared with data in \cite{collaboration_first_2016}. The probabilities of each isotope passing the neutron selection criteria of this analysis were applied to these yields in order to find the expected number of events per muon. Across all cosmogenic isotopes, the total number of expected background events per muon is $\mathcal{O}(10^{-7}$), which is considered to be negligible \cite{liggins_cosmic_2020}.

The other contribution comes from random background events, from internal and external radioactivity, that occur in coincidence with a muon. The rate of events which pass the neutron selection criteria is evaluated over approximately hour-long periods, outside each cosmogenic muon's time window. This rate is then multiplied by the neutron selection time window of $766\,\mu\textup{s}$ to give the number of events expected during the selected event window following each muon. The total number of random background events, across all muons, was found to be $150\pm2(\textup{stat})$.

\subsection{\label{sec:Efficiency}Neutron Selection Efficiency}
The neutron selection efficiency is defined as the fraction of neutrons produced along a section of the muon track within the detector that are detected and pass the neutron selection criteria. Simulation can be used to evaluate this fraction. For each neutron generated in simulation, the probability of selecting that neutron can be calculated, after taking into account reconstruction uncertainties and effects due to the detector response, including trigger efficiency.

Am-Be calibration data has demonstrated that the efficiency for detecting neutrons depends on position within the detector \cite{collaboration_measurement_2020}. For events close to the centre of the detector, light from the neutron capture signal has to travel further and is more susceptible to scattering. Scattering leads to a larger position reconstruction uncertainty, with some events even failing reconstruction. In this region, the efficiency is expected to be lower, and to gradually rise moving away from the centre. This slow rise ends, and the efficiency starts to decrease near the PMTs due to the lower angular acceptance of the PMTs at large incidence angles. The lower angular acceptance of the PMTs leads to a very small number of neutrons being reconstructed at large radii. Moreover, effects due to the fiducial volume start to come into play at large radii. Since each muon track is unique, spanning different lengths and regions of the detector, the neutron selection efficiency must be calculated for each muon in data.

\begin{figure}
    \centering
    \includegraphics[width=\linewidth]{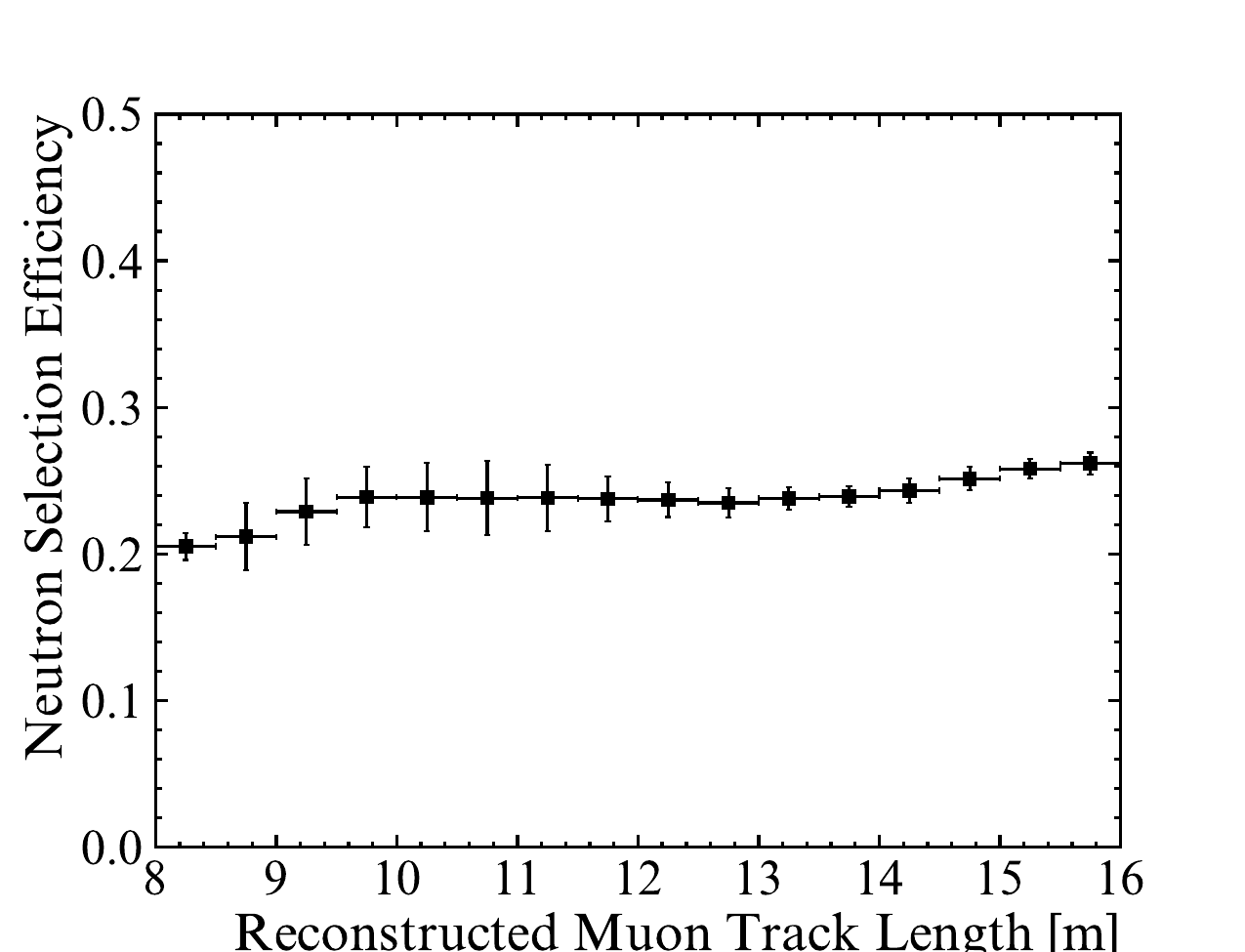}
    \caption{\label{fig:efficiency_fit}The neutron selection efficiencies calculated using $10\,\%$ of the muons in data, binned by the reconstructed muon track length.}
\end{figure}

Figure~\ref{fig:efficiency_fit} shows the neutron selection efficiencies binned according to the reconstructed muon track length, as defined in Section~\ref{sec:track_length}, calculated using $10\,\%$ of the muons selected from the data. The value of each bin is found using the average of all efficiencies within the bin, weighted by their statistical uncertainties. The uncertainty of each bin is given by the spread of the efficiencies within the bin. Each muon in the data is assigned a neutron selection efficiency and uncertainty based on the bin corresponding to its reconstructed track length \cite{dixon_cosmic_2025}.

The neutron selection efficiency is calculated from simulating muons, as described in Section~\ref{sec:Simulation}, where the neutron selection criteria of Section~\ref{sec:Muon_Selection} are applied to the neutrons produced in the simulations. Each muon is simulated using the reconstructed directions and initial positions from the selected muons in data. The muon reconstruction uncertainty is accounted for by displacing both the direction and initial position by a pair of angles sampled from Figure~\ref{fig:muonFitter}.

\begin{figure}
    \centering
    \includegraphics[width=\linewidth]{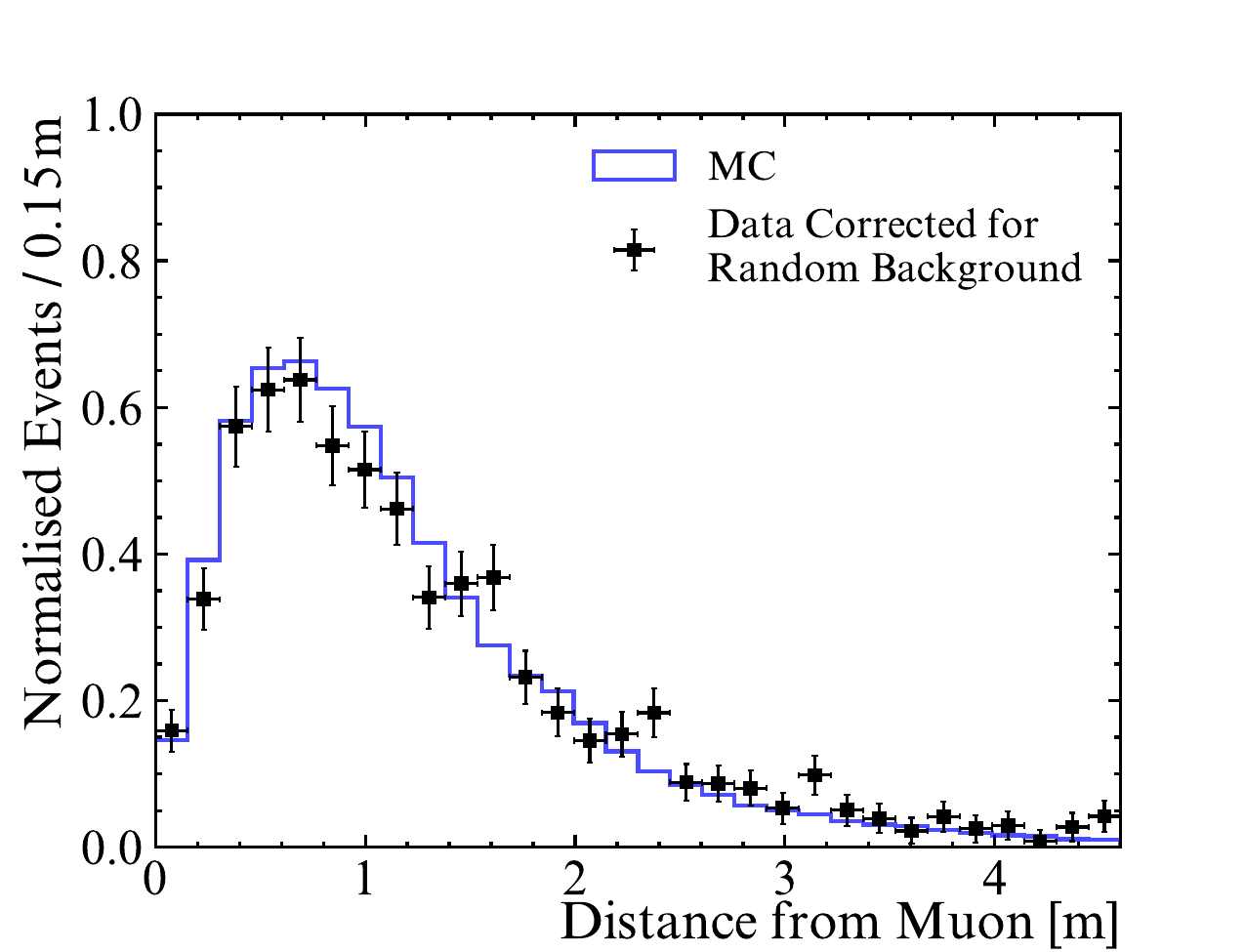}
    \caption{\label{fig:dist_DATAvsMC}The perpendicular distance between the muon track and the neutron capture positions in data (black), compared to the distribution in MC (blue).}
\end{figure}

\begin{figure}
    \centering
    \includegraphics[width=\linewidth]{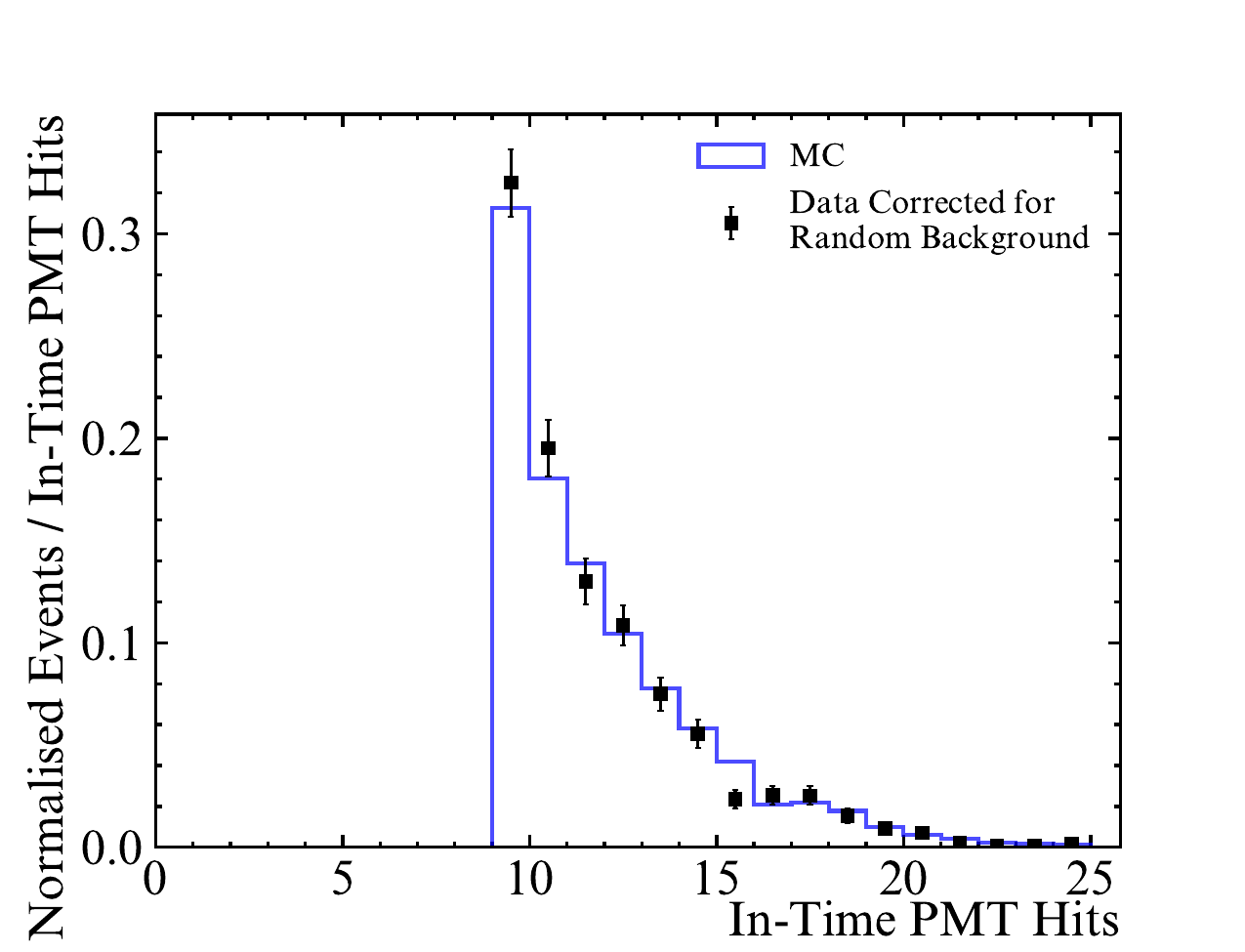}
    \caption{\label{fig:hits_DATAvsMC}The number of in-time PMT hits for the neutron captures in data (black), compared to the distribution in MC (blue).}
\end{figure}

Potential mismodelling of neutron captures in the simulations were also considered. The two relevant observables investigated were: the distance neutrons travel before capture, and the light collected by the detector following capture. Both situations have been constrained through the comparisons between data and MC for the perpendicular distance between the muon track and the neutron capture position, and the number of in-time PMT hits. The direct comparisons of these quantities are shown in Figures~\ref{fig:dist_DATAvsMC}~and~\ref{fig:hits_DATAvsMC}, respectively. The data points in these plots are corrected by subtracting the data-derived expected distributions of random backgrounds.

To investigate the possible impact of the observed data-MC differences in Figures~\ref{fig:dist_DATAvsMC}~and~\ref{fig:hits_DATAvsMC} on the overall neutron selection efficiency, the value of those parameters were scaled event-by-event in the MC, the neutron selection criteria re-applied, and the distributions re-generated. This allowed scaling factors that optimize the data-MC agreement to be determined, and the change in the MC neutron selection efficiency between the scaled and the un-scaled cases to be evaluated. For the perpendicular distance traversed by neutrons, a scale factor of $1.07\pm0.03$ is preferred. Applying this scale factor results in a neutron selection efficiency that is reduced by $4-11\,\%$. This scale factor is referred to as the Distance Correction. For the number of in-time PMT hits, a scale factor of $0.95\pm0.01$ is preferred by the muon-induced neutrons, resulting in a reduction of approximately $14\,\%$ in the neutron selection efficiency. This scale factor is referred to as the Cosmogenic Hits correction. In addition, this method is applied to the number of in-time PMT hits produced by neutron captures from the Am-Be calibration data. Using the Am-Be data significantly increases the number of events. The Am-Be data prefers a scale factor of $0.99\pm0.02$, which causes an approximate $3\,\%$ reduction in the neutron selection efficiency. This scale factor is referred to as the Am-Be Hits correction. Table~\ref{tab:eff-corrections} lists the binned neutron selection efficiencies before corrections, from Figure~\ref{fig:efficiency_fit}, and the resulting neutron selection efficiencies after applying the above corrections. The use of the corrections to the neutron selection efficiency in the final yield calculation will be discussed in Section~\ref{sec:Results}. 
\begin{table*}
\caption{\label{tab:eff-corrections}The neutron selection efficiencies for each muon track length bin, before and  after applying the various corrections, as described in the text.}
\begin{ruledtabular}
\begin{tabular}{ccccc}
\multirow{2}{3cm}{\centering Muon Track\\Length Bin [m]} & \multicolumn{4}{c}{Neutron Selection Efficiency} \\
 & No correction & Distance correction & Am-Be Hits correction & Cosmogenic Hits correction \\
\hline
8.0--8.5  & $0.205 \pm 0.009$ & $0.183 \pm 0.012$ & $0.200 \pm 0.012$ & $0.176 \pm 0.006$ \\
8.5--9.0  & $0.212 \pm 0.023$ & $0.193 \pm 0.020$ & $0.206 \pm 0.012$ & $0.181 \pm 0.006$ \\
9.0--9.5  & $0.229 \pm 0.023$ & $0.207 \pm 0.020$ & $0.223 \pm 0.013$ & $0.196 \pm 0.007$ \\
9.5--10.0 & $0.239 \pm 0.021$ & $0.218 \pm 0.016$ & $0.232 \pm 0.014$ & $0.204 \pm 0.007$ \\
10.0--10.5 & $0.239 \pm 0.023$ & $0.219 \pm 0.020$ & $0.232 \pm 0.014$ & $0.204 \pm 0.007$ \\
10.5--11.0 & $0.238 \pm 0.025$ & $0.217 \pm 0.025$ & $0.232 \pm 0.014$ & $0.204 \pm 0.007$ \\
11.0--11.5 & $0.238 \pm 0.023$ & $0.219 \pm 0.023$ & $0.232 \pm 0.014$ & $0.204 \pm 0.007$ \\
11.5--12.0 & $0.238 \pm 0.016$ & $0.222 \pm 0.013$ & $0.231 \pm 0.014$ & $0.203 \pm 0.007$ \\
12.0--12.5 & $0.237 \pm 0.012$ & $0.220 \pm 0.012$ & $0.231 \pm 0.014$ & $0.203 \pm 0.007$ \\
12.5--13.0 & $0.235 \pm 0.010$ & $0.222 \pm 0.010$ & $0.229 \pm 0.014$ & $0.201 \pm 0.007$ \\
13.0--13.5 & $0.238 \pm 0.008$ & $0.224 \pm 0.009$ & $0.232 \pm 0.014$ & $0.204 \pm 0.007$ \\
13.5--14.0 & $0.239 \pm 0.007$ & $0.227 \pm 0.008$ & $0.233 \pm 0.014$ & $0.205 \pm 0.007$ \\
14.0--14.5 & $0.243 \pm 0.008$ & $0.231 \pm 0.009$ & $0.237 \pm 0.014$ & $0.208 \pm 0.007$ \\
14.5--15.0 & $0.251 \pm 0.008$ & $0.240 \pm 0.009$ & $0.245 \pm 0.015$ & $0.215 \pm 0.007$ \\
15.0--15.5 & $0.258 \pm 0.007$ & $0.248 \pm 0.008$ & $0.251 \pm 0.015$ & $0.221 \pm 0.008$ \\
15.5--16.0 & $0.262 \pm 0.008$ & $0.252 \pm 0.008$ & $0.255 \pm 0.015$ & $0.224 \pm 0.008$ \\
\end{tabular}
\end{ruledtabular}
\end{table*}

\section{\label{sec:Results}Results and Conclusions}
Using Equation~\ref{eq:yield}, the muon-induced neutron yield in water at SNO+ is $Y_{n}=(3.38^{+0.23}_{-0.30})\times10^{-4}\,\textup{cm}^{2}\textup{g}^{-1}\mu^{-1}$. As discussed in Section~\ref{sec:Efficiency}, there are corrections to the neutron selection efficiency that need to be considered. The error range in this result is chosen to cover the range of corrections. The lower error corresponds to the non-corrected neutron selection efficiency, and the upper error corresponds to the Cosmogenic Hits correction. The central value in this result uses the Distance correction, further corrected by the Am-Be Hits correction. This conservative approach to estimating the uncertainty ensures that the uncertainty in the yield accounts for any neutron model dependencies. The $3\,\%$ statistical uncertainty has been included in the final uncertainty.

\begin{figure}
    \centering
    \includegraphics[width=\linewidth]{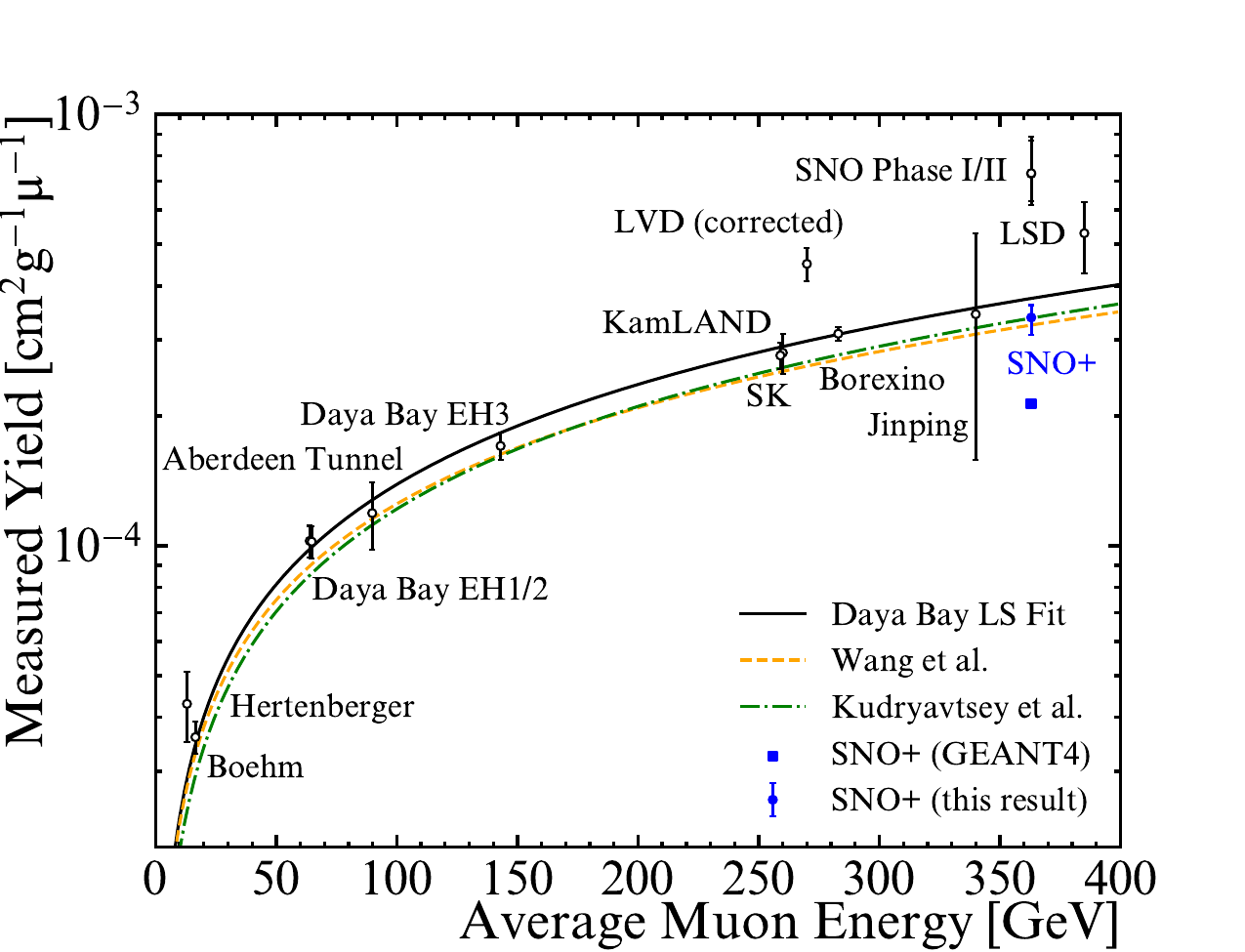}
    \caption{\label{fig:all_yields}The neutron yield measured by different experiments with varying average muon energies. The black empty markers represent measurements made by other experiments \cite{an_cosmogenic_2018,aglietta_neutron_1989,zhao_measurement_2022,abe_production_2010,boehm_neutron_2000,hertenberger_muon-induced_1995, blyth_measurement_2016,bellini_cosmogenic_2013,collaboration_measurement_1999,collaboration_measurement_2023,aharmim_cosmogenic_2019}. The solid black line shows the fit made by Daya Bay using some of the liquid scintillator results \cite{an_cosmogenic_2018}. The blue round marker is the SNO+ result of this paper, and the blue square marker shows the \textsc{GEANT4} predicted result for SNO+. The dashed lines represent the \textsc{FLUKA}-based predictions for liquid scintillator: Wang et al. \cite{wang_predicting_2001} (yellow) and Kudryavtsey et al. \cite{kudryavtsev_simulations_2003} (green).}
\end{figure}

The SNO+ result is compared to the wider experimental field in Figure~\ref{fig:all_yields}. Although the result only agrees with the Daya Bay liquid scintillator global fit to within $1.6\,\sigma$, the result does agree with the \textsc{FLUKA} predictions of the neutron yield as a function of muon energy in scintillator.

The \textsc{GEANT4} prediction for the neutron yield in water at SNO+ has been calculated to be $Y_{n}=(2.130\pm0.001(\textup{stat)})\times10^{-4}\,\textup{cm}^{2}\textup{g}^{-1}\mu^{-1}$. This prediction for the neutron yield is $\sim30\,\%$ less than the result measured. A similar deficit in neutron production between \textsc{FLUKA} and \textsc{GEANT4} of up to $30\,\%$ was observed for muons of energy $\gtrsim100\,\textup{GeV}$ in hydrocarbon materials \cite{araujo_muon-induced_2005}. Daya Bay also reported a similar observation between the two models in liquid scintillator \cite{an_cosmogenic_2018}. We note that the discrepancy appears to be almost entirely due to single neutron events; as can be seen in Figure~\ref{fig:numNeutrons-dataMC}, the \textsc{GEANT4} prediction shows an excess in muons producing only one neutron compared with data. This effect was also observed in \cite{an_cosmogenic_2018}.

Another interesting comparison is between the SNO result and this one. As already stated, these measurements share the same muon flux, but SNO used heavy water as their target medium. In a molecule of water, only oxygen has neutrons that can be released in muon interactions, while molecules of heavy water have two deuterium isotopes that can also release neutrons. The SNO+ result shows a significantly lower neutron yield from the SNO result, which could be explained by the available neutron in the deuterium being liberated by muon interactions, compared to the lack of deuterium in ordinary water in SNO+.

The SNO+ detector is currently filled with liquid scintillator. Repeating this measurement on the scintillator data will allow a comparison between water and scintillator, using the same muon energy. The same comparison was made between SK and KamLAND, both results agreeing with each other. The SNO+ water-scintillator comparison will complement the previous observation. Along with the comparison with SNO, the SNO+ water and scintillator neutron yield measurements will provide greater insight into how nuclear composition affects the muon-induced neutron production.

\begin{acknowledgments}
\input{Acknowledgements}
\end{acknowledgments}

\bibliographystyle{apsrev4-2}
\bibliography{bibliography}

\end{document}

\title{Cosmogenic Neutron Production in Water at SNO+ - Supplemental Material}
\maketitle
This document provides the supplemental material for the paper `Cosmogenic Neutron Production in Water at SNO+' by the SNO+ experiment. The information provided is for the selected muon, neutron and random background data, and is presented as histograms. Each histogram comes with a table detailing the number of events per bin.

\begin{figure}[h]
    \centering
    \includegraphics[width=0.7\linewidth]{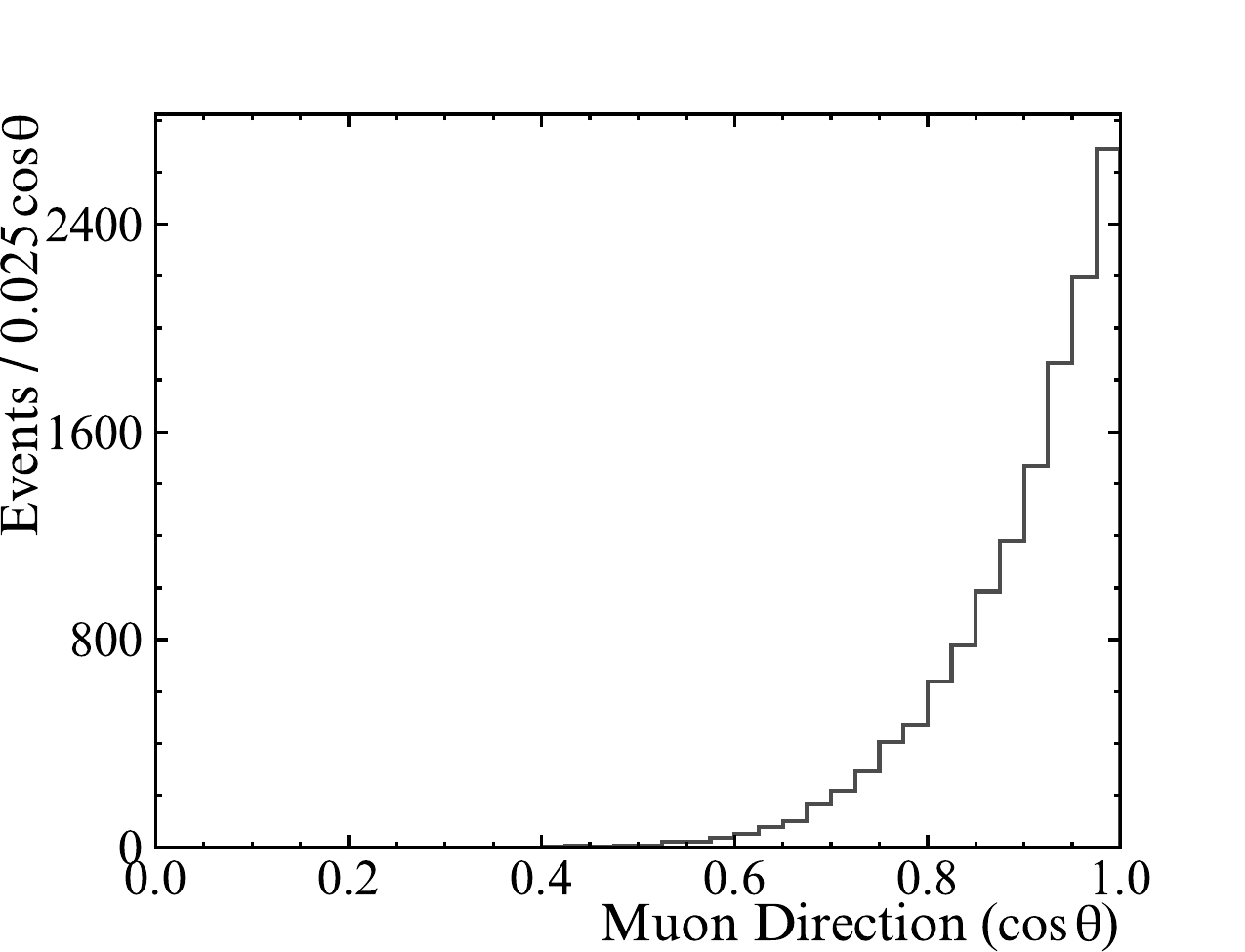}
    \caption{\label{fig:datadirection}The reconstructed zenith angles of the selected cosmic muons' directions. See Table~\ref{tab:data_direction} for bin information.}
\end{figure}
\begin{table}[h]
\caption{\label{tab:data_direction}The number of events per muon direction bin, as presented in Figure~\ref{fig:datadirection}.}
\begin{ruledtabular}
\begin{tabular}{ l c | l c } 
Muon Direction Bin ($\cos{\theta}$) & Events per bin & Muon Direction Bin ($\cos{\theta}$) & Events per bin \\
\hline
$0.40$--$0.425$ & $3$    & $0.70$--$0.725$ & $217$  \\
$0.425$--$0.45$ & $5$    & $0.725$--$0.75$ & $292$  \\
$0.45$--$0.475$ & $3$    & $0.75$--$0.775$ & $405$  \\
$0.475$--$0.50$ & $6$    & $0.775$--$0.80$ & $471$  \\
$0.50$--$0.525$ & $7$    & $0.80$--$0.825$ & $638$  \\
$0.525$--$0.55$ & $22$   & $0.825$--$0.85$ & $779$  \\
$0.55$--$0.575$ & $21$   & $0.85$--$0.875$ & $986$  \\
$0.575$--$0.60$ & $37$   & $0.875$--$0.90$ & $1180$ \\
$0.60$--$0.625$ & $53$   & $0.90$--$0.925$ & $1470$ \\
$0.625$--$0.65$ & $79$   & $0.925$--$0.95$ & $1863$ \\
$0.65$--$0.675$ & $102$  & $0.95$--$0.975$ & $2194$ \\
$0.675$--$0.70$ & $168$  & $0.975$--$1.00$ & $2689$ \\
\end{tabular}
\end{ruledtabular}
\end{table}

\begin{figure}
    \centering
    \includegraphics[width=0.7\linewidth]{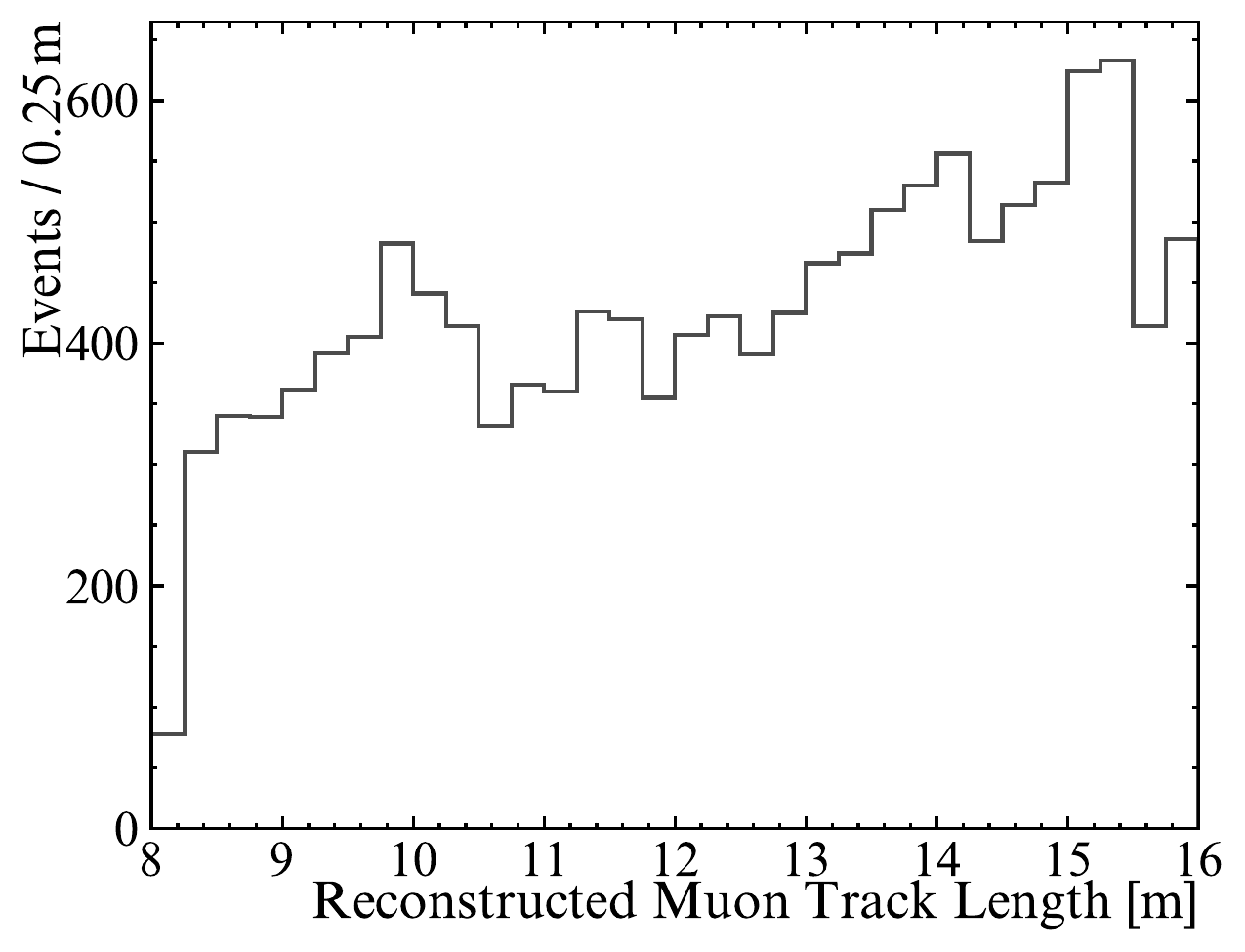}
    \caption{\label{fig:data_path_length}The reconstructed track lengths of the selected cosmic muons. The track length here is defined as the distance between the muon's entry and exit points on the $8\,\textup{m}$ radius sphere, concentric with the PSUP. See Table~\ref{tab:data_path_length} for bin information.}
\end{figure}
\begin{table}[h]
\caption{\label{tab:data_path_length}The number of events per muon track length bin, as presented in Figure~\ref{fig:data_path_length}.}
\begin{ruledtabular}
\begin{tabular}{ l c | l c } 
Muon Track Length Bin [m] & Events per bin & Muon Track Length Bin [m] & Events per bin \\
\hline
$8.00$--$8.25$   & $78$  & $12.00$--$12.25$ & $407$ \\
$8.25$--$8.50$   & $310$ & $12.25$--$12.50$ & $422$ \\
$8.50$--$8.75$   & $340$ & $12.50$--$12.75$ & $391$ \\
$8.75$--$9.00$   & $339$ & $12.75$--$13.00$ & $425$ \\
$9.00$--$9.25$   & $362$ & $13.00$--$13.25$ & $466$ \\
$9.25$--$9.50$   & $392$ & $13.25$--$13.50$ & $474$ \\
$9.50$--$9.75$   & $405$ & $13.50$--$13.75$ & $510$ \\
$9.75$--$10.00$  & $482$ & $13.75$--$14.00$ & $530$ \\
$10.00$--$10.25$ & $441$ & $14.00$--$14.25$ & $556$ \\
$10.25$--$10.50$ & $414$ & $14.25$--$14.50$ & $484$ \\
$10.50$--$10.75$ & $332$ & $14.50$--$14.75$ & $514$ \\
$10.75$--$11.00$ & $366$ & $14.75$--$15.00$ & $532$ \\
$11.00$--$11.25$ & $360$ & $15.00$--$15.25$ & $624$ \\
$11.25$--$11.50$ & $426$ & $15.25$--$15.50$ & $633$ \\
$11.50$--$11.75$ & $420$ & $15.50$--$15.75$ & $414$ \\
$11.75$--$12.00$ & $355$ & $15.75$--$16.00$ & $486$ \\
\end{tabular}
\end{ruledtabular}
\end{table}

\begin{figure}[h]
    \centering
    \includegraphics[width=0.7\linewidth]{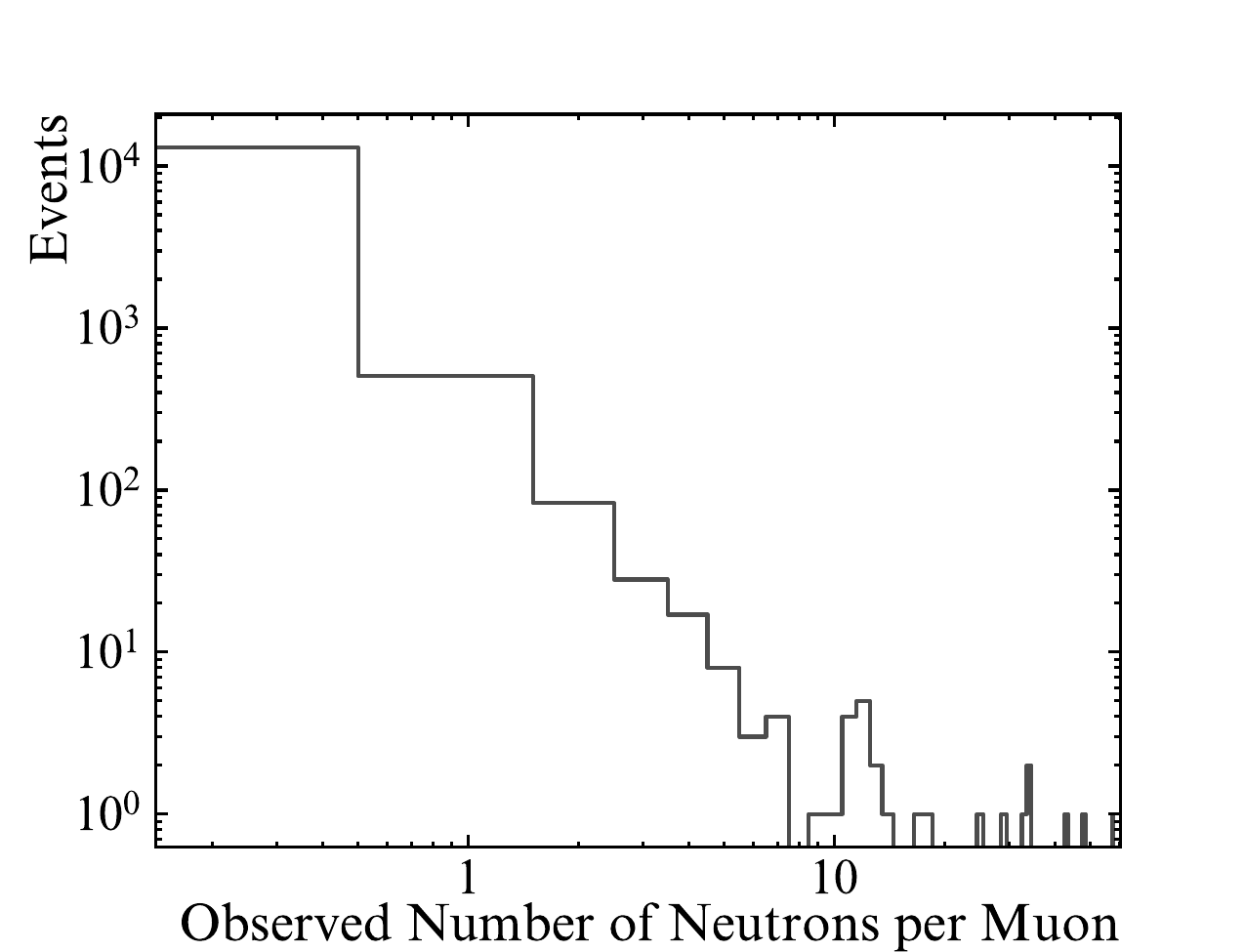}
    \caption{\label{fig:data_obs_neutrons}The number of selected neutrons per cosmic muon. See Table~\ref{tab:data_obs_neutrons} for bin information.}
\end{figure}
\begin{table}[h]
\caption{\label{tab:data_obs_neutrons}The number of events per selected number of neutrons bin, as presented in Figure~\ref{fig:data_obs_neutrons}.}
\begin{ruledtabular}
\begin{tabular}{ l c | l c } 
Number of Neutrons Bin & Events per bin & Number of Neutrons Bin & Events per bin \\
\hline
$0$  & $13018$ & $30$ & $1$ \\
$1$  & $505$   & $31$ & $0$ \\
$2$  & $83$    & $32$ & $0$ \\
$3$  & $28$    & $33$ & $0$ \\
$4$  & $17$    & $34$ & $1$ \\
$5$  & $8$     & $35$ & $2$ \\
$6$  & $3$     & $36$ & $0$ \\
$7$  & $4$     & $37$ & $0$ \\
$8$  & $0$     & $38$ & $0$ \\
$9$  & $1$     & $39$ & $0$ \\
$10$ & $1$     & $40$ & $0$ \\
$11$ & $4$     & $41$ & $0$ \\
$12$ & $5$     & $42$ & $0$ \\
$13$ & $2$     & $43$ & $1$ \\
$14$ & $1$     & $44$ & $0$ \\
$15$ & $0$     & $45$ & $0$ \\
$16$ & $0$     & $46$ & $0$ \\
$17$ & $1$     & $47$ & $0$ \\
$18$ & $1$     & $48$ & $1$ \\
$19$ & $0$     & $49$ & $0$ \\
$20$ & $0$     & $50$ & $0$ \\
$21$ & $0$     & $51$ & $0$ \\
$22$ & $0$     & $52$ & $0$ \\
$23$ & $0$     & $53$ & $0$ \\
$24$ & $0$     & $54$ & $0$ \\
$25$ & $1$     & $55$ & $0$ \\
$26$ & $0$     & $56$ & $0$ \\
$27$ & $0$     & $57$ & $0$ \\
$28$ & $0$     & $58$ & $1$ \\
$29$ & $1$     &      &      \\
\end{tabular}
\end{ruledtabular}
\end{table}

\begin{figure}[h]
    \centering
    \includegraphics[width=0.7\linewidth]{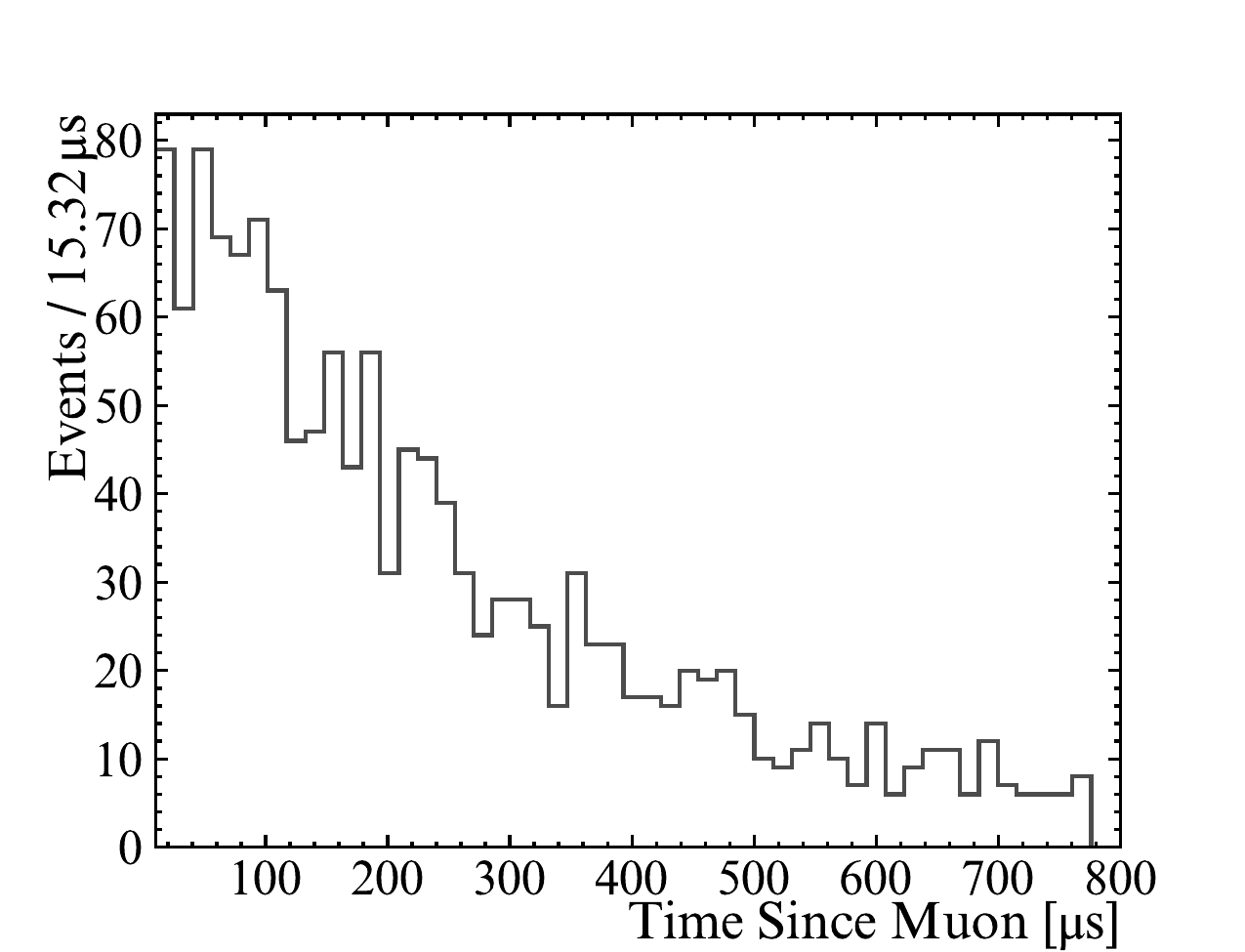}
    \caption{\label{fig:data_neutron_time}The time between selected neutron events and their associated muon event. See Table~\ref{tab:data_neutron_time} for bin information.}
\end{figure}
\begin{table}[h]
\caption{\label{tab:data_neutron_time}The number of events per time since muon bin, as presented in Figure~\ref{fig:data_neutron_time}.}
\begin{ruledtabular}
\begin{tabular}{ l c | l c } 
Time Since Muon Bin [$\mu$s] & Events per bin & Time Since Muon Bin [$\mu$s] & Events per bin \\
\hline
$10.00$--$25.32$ & $79$ & $393.00$--$408.32$ & $17$ \\
$25.32$--$40.64$ & $61$ & $408.32$--$423.64$ & $17$ \\
$40.64$--$55.96$ & $79$ & $423.64$--$438.96$ & $16$ \\
$55.96$--$71.28$ & $69$ & $438.96$--$454.28$ & $20$ \\
$71.28$--$86.60$ & $67$ & $454.28$--$469.60$ & $19$ \\
$86.60$--$101.92$ & $71$ & $469.60$--$484.92$ & $20$ \\
$101.92$--$117.24$ & $63$ & $484.92$--$500.24$ & $15$ \\
$117.24$--$132.56$ & $46$ & $500.24$--$515.56$ & $10$ \\
$132.56$--$147.88$ & $47$ & $515.56$--$530.88$ & $9$ \\
$147.88$--$163.20$ & $56$ & $530.88$--$546.20$ & $11$ \\
$163.20$--$178.52$ & $43$ & $546.20$--$561.52$ & $14$ \\
$178.52$--$193.84$ & $56$ & $561.52$--$576.84$ & $10$ \\
$193.84$--$209.16$ & $31$ & $576.84$--$592.16$ & $7$ \\
$209.16$--$224.48$ & $45$ & $592.16$--$607.48$ & $14$ \\
$224.48$--$239.80$ & $44$ & $607.48$--$622.80$ & $6$ \\
$239.80$--$255.12$ & $39$ & $622.80$--$638.12$ & $9$ \\
$255.12$--$270.44$ & $31$ & $638.12$--$653.44$ & $11$ \\
$270.44$--$285.76$ & $24$ & $653.44$--$668.76$ & $11$ \\
$285.76$--$301.08$ & $28$ & $668.76$--$684.08$ & $6$ \\
$301.08$--$316.40$ & $28$ & $684.08$--$699.40$ & $12$ \\
$316.40$--$331.72$ & $25$ & $699.40$--$714.72$ & $7$ \\
$331.72$--$347.04$ & $16$ & $714.72$--$730.04$ & $6$ \\
$347.04$--$362.36$ & $31$ & $730.04$--$745.36$ & $6$ \\
$362.36$--$377.68$ & $23$ & $745.36$--$760.68$ & $6$ \\
$377.68$--$393.00$ & $23$ & $760.68$--$776.00$ & $8$ \\
\end{tabular}
\end{ruledtabular}
\end{table}

\begin{figure}[h]
    \centering
    \includegraphics[width=0.7\linewidth]{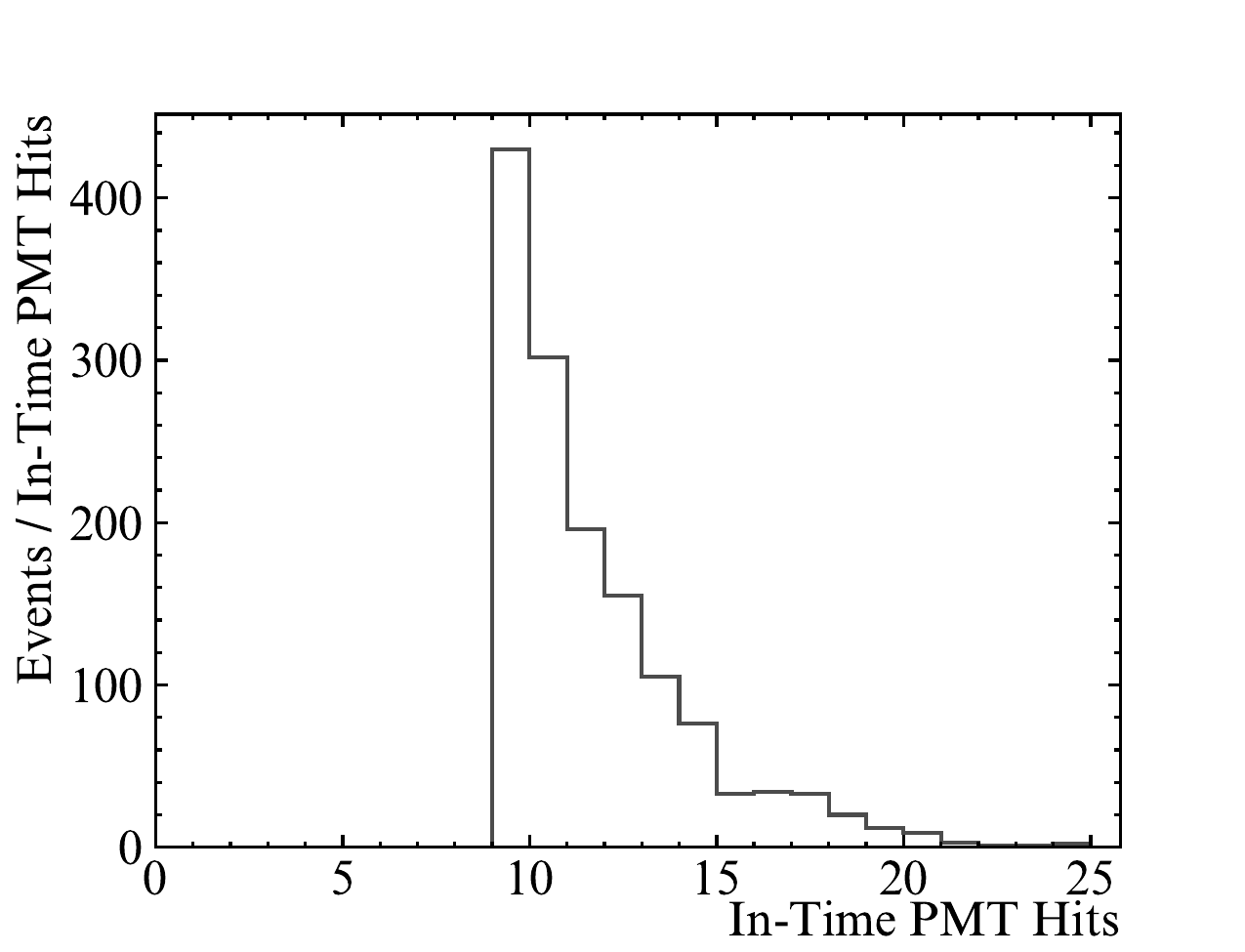}
    \caption{\label{fig:data_neutron_hits}The number of in-time PMT hits of the selected neutron events. See Table~\ref{tab:data_neutron_hits} for bin information.}
\end{figure}
\begin{table}[h]
\caption{\label{tab:data_neutron_hits}The number of events per in-time PMT hits bin, as presented in Figure~\ref{fig:data_neutron_hits}.}
\begin{ruledtabular}
\begin{tabular}{ l c | l c } 
In-Time PMT Hits Bin & Events per bin & In-Time PMT Hits Bin & Events per bin \\
\hline
$9$--$10$   & $430$ & $17$--$18$ & $33$ \\
$10$--$11$  & $302$ & $18$--$19$ & $20$ \\
$11$--$12$  & $196$ & $19$--$20$ & $12$ \\
$12$--$13$  & $155$ & $20$--$21$ & $9$ \\
$13$--$14$  & $105$ & $21$--$22$ & $3$ \\
$14$--$15$  & $76$  & $22$--$23$ & $1$ \\
$15$--$16$  & $33$  & $23$--$24$ & $1$ \\
$16$--$17$  & $34$  & $24$--$25$ & $2$ \\
\end{tabular}
\end{ruledtabular}
\end{table}

\begin{figure}[h]
    \centering
    \includegraphics[width=0.7\linewidth]{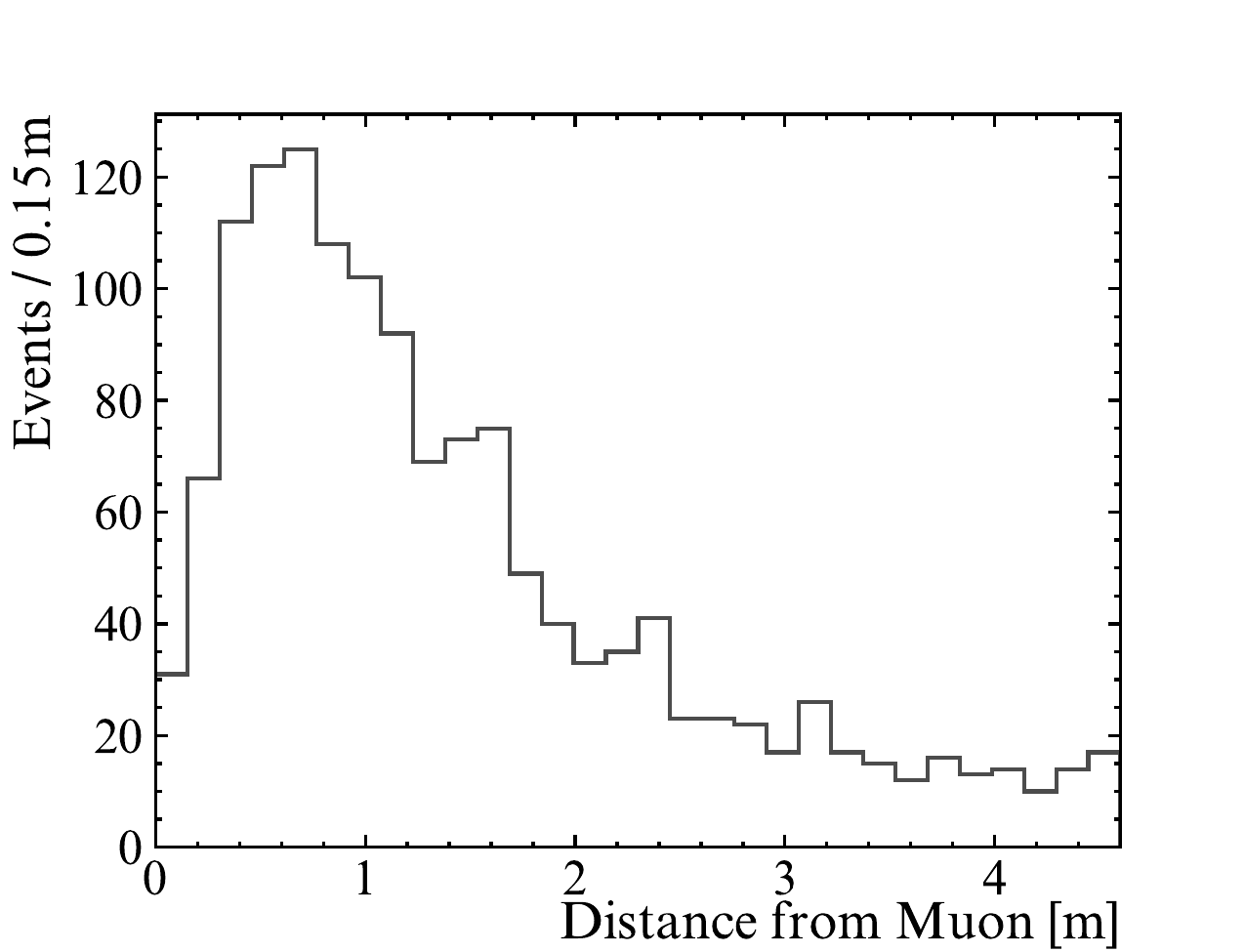}
    \caption{\label{fig:data_neutron_dist}The perpendicular distance between the muon track and the selected neutron capture positions. See Table~\ref{tab:data_neutron_dist} for bin information.}
\end{figure}
\begin{table}[h]
\caption{\label{tab:data_neutron_dist}The number of events per distance from muon bin, as presented in Figure~\ref{fig:data_neutron_dist}.}
\begin{ruledtabular}
\begin{tabular}{ l c | l c } 
Distance from Muon Bin [m] & Events per bin & Distance from Muon Bin [m] & Events per bin \\
\hline
$0.00$--$0.15$ & $31$ & $2.45$--$2.61$ & $23$ \\
$0.15$--$0.31$ & $66$ & $2.61$--$2.76$ & $22$ \\
$0.31$--$0.46$ & $112$ & $2.76$--$2.91$ & $17$ \\
$0.46$--$0.61$ & $122$ & $2.91$--$3.07$ & $26$ \\
$0.61$--$0.77$ & $125$ & $3.07$--$3.22$ & $17$ \\
$0.77$--$0.92$ & $108$ & $3.22$--$3.37$ & $15$ \\
$0.92$--$1.07$ & $102$ & $3.37$--$3.53$ & $12$ \\
$1.07$--$1.23$ & $92$  & $3.53$--$3.68$ & $16$ \\
$1.23$--$1.38$ & $69$  & $3.68$--$3.83$ & $13$ \\
$1.38$--$1.53$ & $73$  & $3.83$--$3.99$ & $14$ \\
$1.53$--$1.69$ & $75$  & $3.99$--$4.14$ & $10$ \\
$1.69$--$1.84$ & $49$  & $4.14$--$4.29$ & $14$ \\
$1.84$--$1.99$ & $40$  & $4.29$--$4.45$ & $17$ \\
$1.99$--$2.15$ & $33$  & $4.45$--$4.60$ & $17$ \\
$2.15$--$2.30$ & $35$  & & \\
\end{tabular}
\end{ruledtabular}
\end{table}

\begin{figure}[h]
    \centering
    \includegraphics[width=0.7\linewidth]{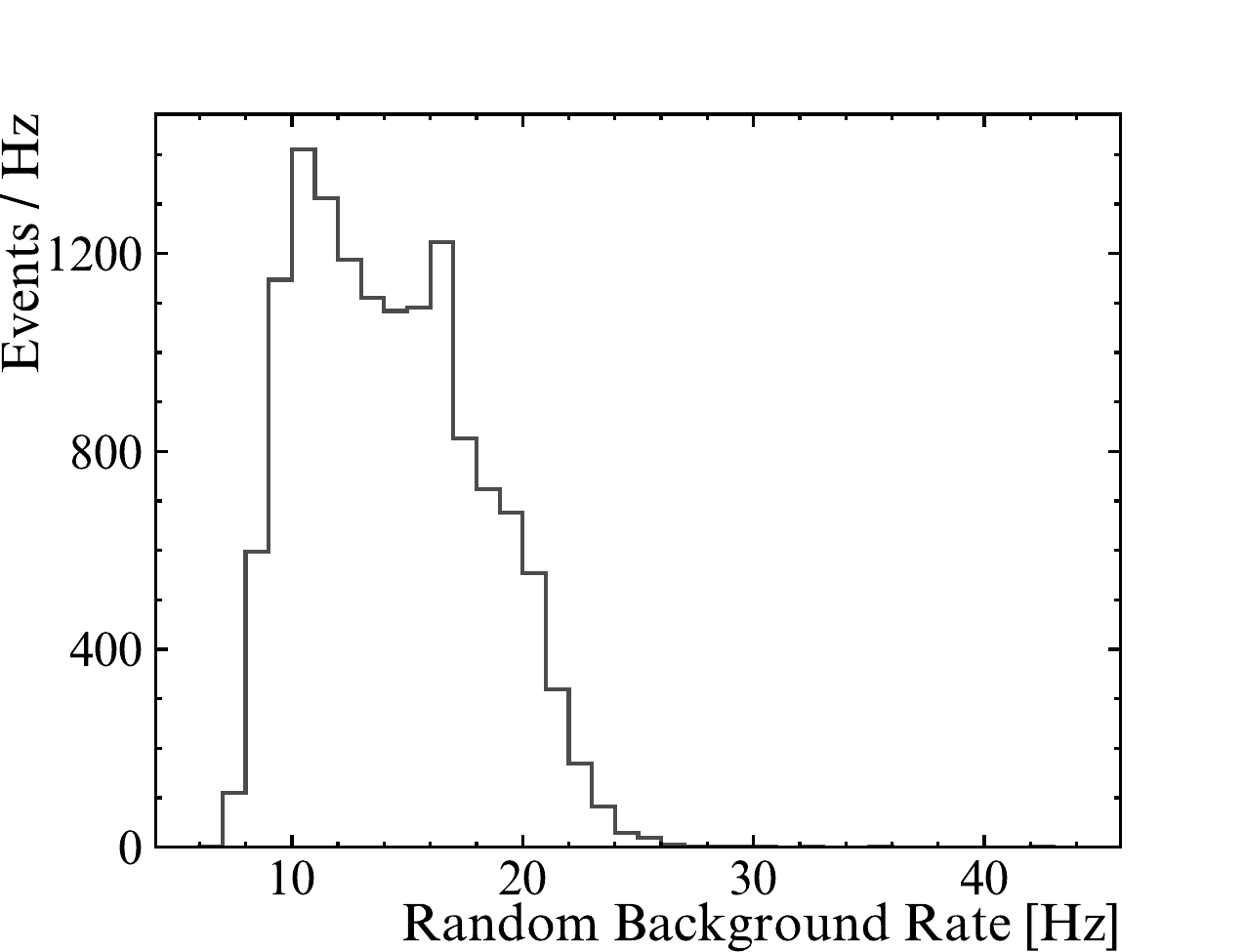}
    \caption{\label{fig:data_random_rate}The random background rate per muon. See Table~\ref{tab:data_random_rate} for bin information.}
\end{figure}
\begin{table}[h]
\caption{\label{tab:data_random_rate}The number of events per random background rate bin, as presented in Figure~\ref{fig:data_random_rate}.}
\begin{ruledtabular}
\begin{tabular}{ l c | l c } 
Rate Bin [Hz] & Events per bin & Rate Bin [Hz] & Events per bin \\
\hline
$6$--$7$   & $2$    & $25$--$26$ & $1$ \\
$7$--$8$   & $110$  & $26$--$27$ & $0$ \\
$8$--$9$   & $597$  & $27$--$28$ & $2$ \\
$9$--$10$  & $1147$ & $28$--$29$ & $0$ \\
$10$--$11$ & $1411$ & $29$--$30$ & $0$ \\
$11$--$12$ & $1312$ & $30$--$31$ & $1$ \\
$12$--$13$ & $1187$ & $31$--$32$ & $0$ \\
$13$--$14$ & $1110$ & $32$--$33$ & $0$ \\
$14$--$15$ & $1084$ & $33$--$34$ & $0$ \\
$15$--$16$ & $1091$ & $34$--$35$ & $0$ \\
$16$--$17$ & $1223$ & $35$--$36$ & $0$ \\
$17$--$18$ & $827$  & $36$--$37$ & $0$ \\
$18$--$19$ & $723$  & $37$--$38$ & $0$ \\
$19$--$20$ & $677$  & $38$--$39$ & $1$ \\
$20$--$21$ & $554$  & $39$--$40$ & $0$ \\
$21$--$22$ & $319$  & $40$--$41$ & $0$ \\
$22$--$23$ & $170$  & $41$--$42$ & $0$ \\
$23$--$24$ & $82$   & $42$--$43$ & $0$ \\
$24$--$25$ & $29$   & $43$--$44$ & $0$ \\
\end{tabular}
\end{ruledtabular}
\end{table}

%% file: authorlist_2025-11-04.tex
% LaTeX Author List
\author{ M.\,Abreu }
\affiliation{ Laborat\'{o}rio de Instrumenta\c{c}\~{a}o e  F\'{\i}sica Experimental de Part\'{\i}culas (LIP), Avenida Professor Gama Pinto, 2, 1649-003, Lisboa, Portugal }
\affiliation{ Universidade de Lisboa, Instituto Superior T\'{e}cnico (IST), Departamento de F\'{\i}sica, Avenida Rovisco Pais, 1049-001 Lisboa, Portugal }
\author{ A.\,Allega }
\affiliation{ Queen's University, Department of Physics, Engineering Physics \& Astronomy, Kingston, Ontario K7L 3N6, Canada }
\author{ M.\,R.\,Anderson }
\affiliation{ Queen's University, Department of Physics, Engineering Physics \& Astronomy, Kingston, Ontario K7L 3N6, Canada }
\author{ S.\,Andringa }
\affiliation{ Laborat\'{o}rio de Instrumenta\c{c}\~{a}o e  F\'{\i}sica Experimental de Part\'{\i}culas (LIP), Avenida Professor Gama Pinto, 2, 1649-003, Lisboa, Portugal }
\author{ D.\,M.\,Asner }
\affiliation{ SNOLAB, Creighton Mine \#9, 1039 Regional Road 24, Sudbury, Ontario P3Y 1N2, Canada }
\author{ D.\,J.\,Auty }
\affiliation{ University of Alberta, Department of Physics, 4-181 CCIS,  Edmonton, Alberta T6G 2E1, Canada }
\author{ A.\,Bacon }
\affiliation{ University of Pennsylvania, Department of Physics \& Astronomy, 209 South 33rd Street, Philadelphia, Pennsylvania 19104-6396, USA }
\author{ T.\,Baltazar }
\affiliation{ Laborat\'{o}rio de Instrumenta\c{c}\~{a}o e  F\'{\i}sica Experimental de Part\'{\i}culas (LIP), Avenida Professor Gama Pinto, 2, 1649-003, Lisboa, Portugal }
\affiliation{ Universidade de Lisboa, Instituto Superior T\'{e}cnico (IST), Departamento de F\'{\i}sica, Avenida Rovisco Pais, 1049-001 Lisboa, Portugal }
\author{ F.\,Bar\~{a}o }
\affiliation{ Laborat\'{o}rio de Instrumenta\c{c}\~{a}o e  F\'{\i}sica Experimental de Part\'{\i}culas (LIP), Avenida Professor Gama Pinto, 2, 1649-003, Lisboa, Portugal }
\affiliation{ Universidade de Lisboa, Instituto Superior T\'{e}cnico (IST), Departamento de F\'{\i}sica, Avenida Rovisco Pais, 1049-001 Lisboa, Portugal }
\author{ N.\,Barros }
\affiliation{ Departamento de F\'{\i}sica, Escola de Ciências, Universidade do Minho, 4710-057 Braga, Portugal}
\affiliation{LIP -- Laborat\'{o}rio de Instrumenta\c{c}\~{a}o e F\'{\i}sica Experimental de Part\'{i}culas, Escola de Ci\^{e}ncias, Campus de Gualtar, Universidade do Minho, 4701-057 Braga, Portugal}
% ...
\collaboration{ The SNO+ Collaboration }

%% file: Acknowledgements.tex
Capital funds for SNO+ were provided by the Canada Foundation for Innovation and matching partners: Ontario Ministry of Research, Innovation and Science, Alberta Science and Research Investments Program, Queen's University at Kingston, and the Federal Economic Development Agency for Northern Ontario. This research was supported by (Canada) the Natural Sciences and Engineering Research Council of Canada, the Canadian Institute for Advanced Research, the Ontario Early Researcher Awards; (U.S.) the Department of Energy (DOE) Office of Nuclear Physics, the National Science Foundation, and the DOE National Nuclear Security Administration through the Nuclear Science and Security Consortium; (U.K.) the Science and Technology Facilities Council, the Royal Society, and an NMES studentship from King's College London; (Portugal) Fundação para a Ciência e a Tecnologia (FCT-Portugal); (Germany) the Deutsche Forschungsgemeinschaft; (Mexico) DGAPA-UNAM and Consejo Nacional de Ciencia y Tecnología; and (China) the Discipline Construction Fund of Shandong University. We also thank SNOLAB and SNO+ technical staff; the Digital Research Alliance of Canada; the GridPP Collaboration and support from Rutherford Appleton Laboratory, the King's College London Computational Research, Engineering and Technology Environment (CREATE), and the Savio computational cluster at the University of California, Berkeley. Additional long-term storage was provided by the Fermilab Scientific Computing Division. 

For the purposes of open access, the authors have applied a Creative Commons Attribution licence to any Author Accepted Manuscript version arising. Representations of the data relevant to the conclusions drawn here are provided within this paper and its supplemental material \cite{supp_mat}.

%% file: main.bbl
\begin{thebibliography}{34}%
\makeatletter
\providecommand \@ifxundefined [1]{%
 \@ifx{#1\undefined}
}%
\providecommand \@ifnum [1]{%
 \ifnum #1\expandafter \@firstoftwo
 \else \expandafter \@secondoftwo
 \fi
}%
\providecommand \@ifx [1]{%
 \ifx #1\expandafter \@firstoftwo
 \else \expandafter \@secondoftwo
 \fi
}%
\providecommand \natexlab [1]{#1}%
\providecommand \enquote  [1]{``#1''}%
\providecommand \bibnamefont  [1]{#1}%
\providecommand \bibfnamefont [1]{#1}%
\providecommand \citenamefont [1]{#1}%
\providecommand \href@noop [0]{\@secondoftwo}%
\providecommand \href [0]{\begingroup \@sanitize@url \@href}%
\providecommand \@href[1]{\@@startlink{#1}\@@href}%
\providecommand \@@href[1]{\endgroup#1\@@endlink}%
\providecommand \@sanitize@url [0]{\catcode `\\12\catcode `\$12\catcode `\&12\catcode `\#12\catcode `\^12\catcode `\_12\catcode `\%12\relax}%
\providecommand \@@startlink[1]{}%
\providecommand \@@endlink[0]{}%
\providecommand \url  [0]{\begingroup\@sanitize@url \@url }%
\providecommand \@url [1]{\endgroup\@href {#1}{\urlprefix }}%
\providecommand \urlprefix  [0]{URL }%
\providecommand \Eprint [0]{\href }%
\providecommand \doibase [0]{https://doi.org/}%
\providecommand \selectlanguage [0]{\@gobble}%
\providecommand \bibinfo  [0]{\@secondoftwo}%
\providecommand \bibfield  [0]{\@secondoftwo}%
\providecommand \translation [1]{[#1]}%
\providecommand \BibitemOpen [0]{}%
\providecommand \bibitemStop [0]{}%
\providecommand \bibitemNoStop [0]{.\EOS\space}%
\providecommand \EOS [0]{\spacefactor3000\relax}%
\providecommand \BibitemShut  [1]{\csname bibitem#1\endcsname}%
\let\auto@bib@innerbib\@empty
%</preamble>
\bibitem [{\citenamefont {Wang}\ \emph {et~al.}(2001)\citenamefont {Wang}, \citenamefont {Balic}, \citenamefont {Gratta}, \citenamefont {Fassò}, \citenamefont {Roesler},\ and\ \citenamefont {Ferrari}}]{wang_predicting_2001}%
  \BibitemOpen
  \bibfield  {author} {\bibinfo {author} {\bibfnamefont {Y.-F.}\ \bibnamefont {Wang}}, \bibinfo {author} {\bibfnamefont {V.}~\bibnamefont {Balic}}, \bibinfo {author} {\bibfnamefont {G.}~\bibnamefont {Gratta}}, \bibinfo {author} {\bibfnamefont {A.}~\bibnamefont {Fassò}}, \bibinfo {author} {\bibfnamefont {S.}~\bibnamefont {Roesler}},\ and\ \bibinfo {author} {\bibfnamefont {A.}~\bibnamefont {Ferrari}},\ }\href {https://link.aps.org/doi/10.1103/PhysRevD.64.013012} {\bibfield  {journal} {\bibinfo  {journal} {Phys. Rev. D}\ }\textbf {\bibinfo {volume} {64}},\ \bibinfo {pages} {013012} (\bibinfo {year} {2001})}\BibitemShut {NoStop}%
\bibitem [{\citenamefont {Nairat}\ \emph {et~al.}(2025)\citenamefont {Nairat}, \citenamefont {Beacom},\ and\ \citenamefont {Li}}]{nairat_neutron_2024}%
  \BibitemOpen
  \bibfield  {author} {\bibinfo {author} {\bibfnamefont {O.}~\bibnamefont {Nairat}}, \bibinfo {author} {\bibfnamefont {J.~F.}\ \bibnamefont {Beacom}},\ and\ \bibinfo {author} {\bibfnamefont {S.~W.}\ \bibnamefont {Li}},\ }\href {https://link.aps.org/doi/10.1103/PhysRevD.111.023014} {\bibfield  {journal} {\bibinfo  {journal} {Phys. Rev. D}\ }\textbf {\bibinfo {volume} {111}},\ \bibinfo {pages} {023014} (\bibinfo {year} {2025})}\BibitemShut {NoStop}%
\bibitem [{\citenamefont {An}\ \emph {et~al.}(2018)\citenamefont {An} \emph {et~al.}}]{an_cosmogenic_2018}%
  \BibitemOpen
  \bibfield  {author} {\bibinfo {author} {\bibfnamefont {F.~P.}\ \bibnamefont {An}} \emph {et~al.} (\bibinfo {collaboration} {Daya Bay Collaboration}),\ }\href {https://link.aps.org/doi/10.1103/PhysRevD.97.052009} {\bibfield  {journal} {\bibinfo  {journal} {Phys. Rev. D}\ }\textbf {\bibinfo {volume} {97}},\ \bibinfo {pages} {052009} (\bibinfo {year} {2018})}\BibitemShut {NoStop}%
\bibitem [{\citenamefont {Hertenberger}\ \emph {et~al.}(1995)\citenamefont {Hertenberger}, \citenamefont {Chen},\ and\ \citenamefont {Dougherty}}]{hertenberger_muon-induced_1995}%
  \BibitemOpen
  \bibfield  {author} {\bibinfo {author} {\bibfnamefont {R.}~\bibnamefont {Hertenberger}}, \bibinfo {author} {\bibfnamefont {M.}~\bibnamefont {Chen}},\ and\ \bibinfo {author} {\bibfnamefont {B.~L.}\ \bibnamefont {Dougherty}},\ }\href {https://link.aps.org/doi/10.1103/PhysRevC.52.3449} {\bibfield  {journal} {\bibinfo  {journal} {Phys. Rev. C}\ }\textbf {\bibinfo {volume} {52}},\ \bibinfo {pages} {3449} (\bibinfo {year} {1995})}\BibitemShut {NoStop}%
\bibitem [{\citenamefont {Boehm}\ \emph {et~al.}(2000)\citenamefont {Boehm} \emph {et~al.}}]{boehm_neutron_2000}%
  \BibitemOpen
  \bibfield  {author} {\bibinfo {author} {\bibfnamefont {F.}~\bibnamefont {Boehm}} \emph {et~al.},\ }\href {https://link.aps.org/doi/10.1103/PhysRevD.62.092005} {\bibfield  {journal} {\bibinfo  {journal} {Phys. Rev. D}\ }\textbf {\bibinfo {volume} {62}},\ \bibinfo {pages} {092005} (\bibinfo {year} {2000})}\BibitemShut {NoStop}%
\bibitem [{\citenamefont {Blyth}\ \emph {et~al.}(2016)\citenamefont {Blyth} \emph {et~al.}}]{blyth_measurement_2016}%
  \BibitemOpen
  \bibfield  {author} {\bibinfo {author} {\bibfnamefont {S.~C.}\ \bibnamefont {Blyth}} \emph {et~al.} (\bibinfo {collaboration} {Aberdeen Tunnel Experiment Collaboration}),\ }\href {https://link.aps.org/doi/10.1103/PhysRevD.93.072005} {\bibfield  {journal} {\bibinfo  {journal} {Phys. Rev. D}\ }\textbf {\bibinfo {volume} {93}},\ \bibinfo {pages} {072005} (\bibinfo {year} {2016})}\BibitemShut {NoStop}%
\bibitem [{\citenamefont {Abe}\ \emph {et~al.}(2010)\citenamefont {Abe} \emph {et~al.}}]{abe_production_2010}%
  \BibitemOpen
  \bibfield  {author} {\bibinfo {author} {\bibfnamefont {S.}~\bibnamefont {Abe}} \emph {et~al.} (\bibinfo {collaboration} {KamLAND Collaboration}),\ }\href {https://link.aps.org/doi/10.1103/PhysRevC.81.025807} {\bibfield  {journal} {\bibinfo  {journal} {Phys. Rev. C}\ }\textbf {\bibinfo {volume} {81}},\ \bibinfo {pages} {025807} (\bibinfo {year} {2010})}\BibitemShut {NoStop}%
\bibitem [{\citenamefont {Aglietta}\ \emph {et~al.}(1999)\citenamefont {Aglietta} \emph {et~al.}}]{collaboration_measurement_1999}%
  \BibitemOpen
  \bibfield  {author} {\bibinfo {author} {\bibfnamefont {M.}~\bibnamefont {Aglietta}} \emph {et~al.} (\bibinfo {collaboration} {LVD Collaboration}),\ }\href {http://arxiv.org/abs/hep-ex/9905047} {\  (\bibinfo {year} {1999})},\ \Eprint {https://arxiv.org/abs/hep-ex/9905047} {arXiv:hep-ex/9905047 [hep-ex]} \BibitemShut {NoStop}%
\bibitem [{\citenamefont {Bellini}\ \emph {et~al.}(2013)\citenamefont {Bellini} \emph {et~al.}}]{bellini_cosmogenic_2013}%
  \BibitemOpen
  \bibfield  {author} {\bibinfo {author} {\bibfnamefont {G.}~\bibnamefont {Bellini}} \emph {et~al.} (\bibinfo {collaboration} {Borexino Collaboration}),\ }\href {https://dx.doi.org/10.1088/1475-7516/2013/08/049} {\bibfield  {journal} {\bibinfo  {journal} {J. Cosmol. Astropart. Phys.}\ }\textbf {\bibinfo {volume} {2013}},\ \bibinfo {pages} {049}}\BibitemShut {NoStop}%
\bibitem [{\citenamefont {Aglietta}\ \emph {et~al.}(1989)\citenamefont {Aglietta} \emph {et~al.}}]{aglietta_neutron_1989}%
  \BibitemOpen
  \bibfield  {author} {\bibinfo {author} {\bibfnamefont {M.}~\bibnamefont {Aglietta}} \emph {et~al.},\ }\href {https://doi.org/10.1007/BF02525079} {\bibfield  {journal} {\bibinfo  {journal} {Il Nuovo Cimento C}\ }\textbf {\bibinfo {volume} {12}},\ \bibinfo {pages} {467} (\bibinfo {year} {1989})}\BibitemShut {NoStop}%
\bibitem [{\citenamefont {Aharmim}\ \emph {et~al.}(2019)\citenamefont {Aharmim} \emph {et~al.}}]{aharmim_cosmogenic_2019}%
  \BibitemOpen
  \bibfield  {author} {\bibinfo {author} {\bibfnamefont {B.}~\bibnamefont {Aharmim}} \emph {et~al.} (\bibinfo {collaboration} {SNO Collaboration}),\ }\href {http://arxiv.org/abs/1909.11728} {\bibfield  {journal} {\bibinfo  {journal} {Phys. Rev. D}\ }\textbf {\bibinfo {volume} {100}},\ \bibinfo {pages} {112005} (\bibinfo {year} {2019})}\BibitemShut {NoStop}%
\bibitem [{\citenamefont {Zhao}\ \emph {et~al.}(2022)\citenamefont {Zhao} \emph {et~al.}}]{zhao_measurement_2022}%
  \BibitemOpen
  \bibfield  {author} {\bibinfo {author} {\bibfnamefont {L.}~\bibnamefont {Zhao}} \emph {et~al.},\ }\href {http://arxiv.org/abs/2108.04010} {\bibfield  {journal} {\bibinfo  {journal} {Chinese Phys. C}\ }\textbf {\bibinfo {volume} {46}},\ \bibinfo {pages} {085001} (\bibinfo {year} {2022})}\BibitemShut {NoStop}%
\bibitem [{\citenamefont {Shinoki}\ \emph {et~al.}(2023)\citenamefont {Shinoki} \emph {et~al.}}]{collaboration_measurement_2023}%
  \BibitemOpen
  \bibfield  {author} {\bibinfo {author} {\bibfnamefont {M.}~\bibnamefont {Shinoki}} \emph {et~al.} (\bibinfo {collaboration} {SK Collaboration}),\ }\href {https://link.aps.org/doi/10.1103/PhysRevD.107.092009} {\bibfield  {journal} {\bibinfo  {journal} {Phys. Rev. D}\ }\textbf {\bibinfo {volume} {107}},\ \bibinfo {pages} {092009} (\bibinfo {year} {2023})}\BibitemShut {NoStop}%
\bibitem [{\citenamefont {Anderson}\ \emph {et~al.}(2019{\natexlab{a}})\citenamefont {Anderson} \emph {et~al.}}]{collaboration_measurement_2019}%
  \BibitemOpen
  \bibfield  {author} {\bibinfo {author} {\bibfnamefont {M.~R.}\ \bibnamefont {Anderson}} \emph {et~al.} (\bibinfo {collaboration} {SNO+ Collaboration}),\ }\href {http://arxiv.org/abs/1812.03355} {\bibfield  {journal} {\bibinfo  {journal} {Phys. Rev. D}\ }\textbf {\bibinfo {volume} {99}},\ \bibinfo {pages} {012012} (\bibinfo {year} {2019}{\natexlab{a}})}\BibitemShut {NoStop}%
\bibitem [{\citenamefont {Allega}\ \emph {et~al.}(2024)\citenamefont {Allega} \emph {et~al.}}]{collaboration_measurement_2024}%
  \BibitemOpen
  \bibfield  {author} {\bibinfo {author} {\bibfnamefont {A.}~\bibnamefont {Allega}} \emph {et~al.} (\bibinfo {collaboration} {SNO+ Collaboration}),\ }\href {http://arxiv.org/abs/2407.17595} {\bibfield  {journal} {\bibinfo  {journal} {Phys. Rev. D}\ }\textbf {\bibinfo {volume} {110}},\ \bibinfo {pages} {122003} (\bibinfo {year} {2024})}\BibitemShut {NoStop}%
\bibitem [{\citenamefont {Anderson}\ \emph {et~al.}(2019{\natexlab{b}})\citenamefont {Anderson} \emph {et~al.}}]{Anderson_2019}%
  \BibitemOpen
  \bibfield  {author} {\bibinfo {author} {\bibfnamefont {M.}~\bibnamefont {Anderson}} \emph {et~al.} (\bibinfo {collaboration} {SNO+ Collaboration}),\ }\href {http://dx.doi.org/10.1103/PhysRevD.99.032008} {\bibfield  {journal} {\bibinfo  {journal} {Phys. Rev. D}\ }\textbf {\bibinfo {volume} {99}} (\bibinfo {year} {2019}{\natexlab{b}})}\BibitemShut {NoStop}%
\bibitem [{\citenamefont {Allega}\ \emph {et~al.}(2022)\citenamefont {Allega} \emph {et~al.}}]{collaboration_improved_2022}%
  \BibitemOpen
  \bibfield  {author} {\bibinfo {author} {\bibfnamefont {A.}~\bibnamefont {Allega}} \emph {et~al.} (\bibinfo {collaboration} {SNO+ Collaboration}),\ }\href {http://arxiv.org/abs/2205.06400} {\bibfield  {journal} {\bibinfo  {journal} {Phys. Rev. D}\ }\textbf {\bibinfo {volume} {105}},\ \bibinfo {pages} {112012} (\bibinfo {year} {2022})}\BibitemShut {NoStop}%
\bibitem [{\citenamefont {Allega}\ \emph {et~al.}(2023)\citenamefont {Allega} \emph {et~al.}}]{collaboration_evidence_2023}%
  \BibitemOpen
  \bibfield  {author} {\bibinfo {author} {\bibfnamefont {A.}~\bibnamefont {Allega}} \emph {et~al.} (\bibinfo {collaboration} {SNO+ Collaboration}),\ }\href {http://arxiv.org/abs/2210.14154} {\bibfield  {journal} {\bibinfo  {journal} {Phys. Rev. Lett.}\ }\textbf {\bibinfo {volume} {130}},\ \bibinfo {pages} {091801} (\bibinfo {year} {2023})}\BibitemShut {NoStop}%
\bibitem [{\citenamefont {Lawson}\ \emph {et~al.}(2013)\citenamefont {Lawson}, \citenamefont {Smith},\ and\ \citenamefont {Jauregui}}]{lawson_snolab_2013}%
  \BibitemOpen
  \bibfield  {author} {\bibinfo {author} {\bibfnamefont {I.}~\bibnamefont {Lawson}}, \bibinfo {author} {\bibfnamefont {N.}~\bibnamefont {Smith}},\ and\ \bibinfo {author} {\bibfnamefont {E.~V.}\ \bibnamefont {Jauregui}},\ }\href {https://doi.org/10.1080/10619127.2013.767692} {\bibfield  {journal} {\bibinfo  {journal} {Nuclear Physics News}\ }\textbf {\bibinfo {volume} {23}},\ \bibinfo {pages} {5} (\bibinfo {year} {2013})}\BibitemShut {NoStop}%
\bibitem [{\citenamefont {Smith}(2012)}]{smith_snolab_2012}%
  \BibitemOpen
  \bibfield  {author} {\bibinfo {author} {\bibfnamefont {N.~J.~T.}\ \bibnamefont {Smith}},\ }\href {https://doi.org/10.1140/epjp/i2012-12108-9} {\bibfield  {journal} {\bibinfo  {journal} {Eur. Phys. J. Plus}\ }\textbf {\bibinfo {volume} {127}},\ \bibinfo {pages} {108} (\bibinfo {year} {2012})}\BibitemShut {NoStop}%
\bibitem [{\citenamefont {Albanese}\ \emph {et~al.}(2021)\citenamefont {Albanese} \emph {et~al.}}]{collaboration_sno_2021}%
  \BibitemOpen
  \bibfield  {author} {\bibinfo {author} {\bibfnamefont {V.}~\bibnamefont {Albanese}} \emph {et~al.} (\bibinfo {collaboration} {SNO+ Collaboration}),\ }\href {http://arxiv.org/abs/2104.11687} {\bibfield  {journal} {\bibinfo  {journal} {JINST}\ }\textbf {\bibinfo {volume} {16}},\ \bibinfo {pages} {P08059}}\BibitemShut {NoStop}%
\bibitem [{\citenamefont {Anderson}\ \emph {et~al.}(2020)\citenamefont {Anderson} \emph {et~al.}}]{collaboration_measurement_2020}%
  \BibitemOpen
  \bibfield  {author} {\bibinfo {author} {\bibfnamefont {M.~R.}\ \bibnamefont {Anderson}} \emph {et~al.} (\bibinfo {collaboration} {SNO+ Collaboration}),\ }\href {http://arxiv.org/abs/2002.10351} {\bibfield  {journal} {\bibinfo  {journal} {Phys. Rev. C}\ }\textbf {\bibinfo {volume} {102}},\ \bibinfo {pages} {014002} (\bibinfo {year} {2020})}\BibitemShut {NoStop}%
\bibitem [{noa()}]{noauthor_rat_nodate}%
  \BibitemOpen
  \href {https://rat.readthedocs.io/en/latest/index.html} {\bibinfo {title} {{RAT} (is an {Analysis} {Tool}) {User}’s {Guide} — {RAT} 1.0 documentation}}\BibitemShut {NoStop}%
\bibitem [{\citenamefont {Agostinelli}\ \emph {et~al.}(2003)\citenamefont {Agostinelli} \emph {et~al.}}]{agostinelli_geant4simulation_2003}%
  \BibitemOpen
  \bibfield  {author} {\bibinfo {author} {\bibfnamefont {S.}~\bibnamefont {Agostinelli}} \emph {et~al.} (\bibinfo {collaboration} {GEANT4 Collaboration}),\ }\href {https://www.sciencedirect.com/science/article/pii/S0168900203013688} {\bibfield  {journal} {\bibinfo  {journal} {Nucl. Instrum. Methods Phys. Res., Sect. A}\ }\textbf {\bibinfo {volume} {506}},\ \bibinfo {pages} {250} (\bibinfo {year} {2003})}\BibitemShut {NoStop}%
\bibitem [{\citenamefont {Mei}\ and\ \citenamefont {Hime}(2006)}]{mei_muon-induced_2006}%
  \BibitemOpen
  \bibfield  {author} {\bibinfo {author} {\bibfnamefont {D.-M.}\ \bibnamefont {Mei}}\ and\ \bibinfo {author} {\bibfnamefont {A.}~\bibnamefont {Hime}},\ }\href {http://arxiv.org/abs/astro-ph/0512125} {\bibfield  {journal} {\bibinfo  {journal} {Phys. Rev. D}\ }\textbf {\bibinfo {volume} {73}},\ \bibinfo {pages} {053004} (\bibinfo {year} {2006})}\BibitemShut {NoStop}%
\bibitem [{\citenamefont {Groom}\ \emph {et~al.}(2001)\citenamefont {Groom}, \citenamefont {Mokhov},\ and\ \citenamefont {Striganov}}]{groom_muon_2001}%
  \BibitemOpen
  \bibfield  {author} {\bibinfo {author} {\bibfnamefont {D.~E.}\ \bibnamefont {Groom}}, \bibinfo {author} {\bibfnamefont {N.~V.}\ \bibnamefont {Mokhov}},\ and\ \bibinfo {author} {\bibfnamefont {S.~I.}\ \bibnamefont {Striganov}},\ }\href {https://doi.org/10.1006/adnd.2001.0861} {\bibfield  {journal} {\bibinfo  {journal} {At. Data Nucl. Data Tables}\ }\textbf {\bibinfo {volume} {78}},\ \bibinfo {pages} {183} (\bibinfo {year} {2001})}\BibitemShut {NoStop}%
\bibitem [{\citenamefont {Liggins}(2020)}]{liggins_cosmic_2020}%
  \BibitemOpen
  \bibfield  {author} {\bibinfo {author} {\bibfnamefont {B.}~\bibnamefont {Liggins}},\ }\emph {\bibinfo {title} {Cosmic muon-induced neutrons in the {SNO}+ water phase}},\ \href {https://qmro.qmul.ac.uk/xmlui/handle/123456789/68614} {Ph.D. thesis},\ \bibinfo  {school} {Queen Mary University of London} (\bibinfo {year} {2020})\BibitemShut {NoStop}%
\bibitem [{\citenamefont {Aharmim}\ \emph {et~al.}(2009)\citenamefont {Aharmim} \emph {et~al.}}]{collaboration_measurement_2009}%
  \BibitemOpen
  \bibfield  {author} {\bibinfo {author} {\bibfnamefont {B.}~\bibnamefont {Aharmim}} \emph {et~al.} (\bibinfo {collaboration} {SNO Collaboration}),\ }\href {http://arxiv.org/abs/0902.2776} {\bibfield  {journal} {\bibinfo  {journal} {Phys. Rev. D}\ }\textbf {\bibinfo {volume} {80}},\ \bibinfo {pages} {012001} (\bibinfo {year} {2009})}\BibitemShut {NoStop}%
\bibitem [{\citenamefont {Li}\ and\ \citenamefont {Beacom}(2014)}]{li_first_2014}%
  \BibitemOpen
  \bibfield  {author} {\bibinfo {author} {\bibfnamefont {S.~W.}\ \bibnamefont {Li}}\ and\ \bibinfo {author} {\bibfnamefont {J.~F.}\ \bibnamefont {Beacom}},\ }\href {http://arxiv.org/abs/1402.4687} {\bibfield  {journal} {\bibinfo  {journal} {Phys. Rev. C}\ }\textbf {\bibinfo {volume} {89}} (\bibinfo {year} {2014})}\BibitemShut {NoStop}%
\bibitem [{\citenamefont {Zhang}\ \emph {et~al.}(2016)\citenamefont {Zhang} \emph {et~al.}}]{collaboration_first_2016}%
  \BibitemOpen
  \bibfield  {author} {\bibinfo {author} {\bibfnamefont {Y.}~\bibnamefont {Zhang}} \emph {et~al.} (\bibinfo {collaboration} {SK Collaboration}),\ }\href {http://arxiv.org/abs/1509.08168} {\bibfield  {journal} {\bibinfo  {journal} {Phys. Rev. D}\ }\textbf {\bibinfo {volume} {93}},\ \bibinfo {pages} {012004} (\bibinfo {year} {2016})}\BibitemShut {NoStop}%
\bibitem [{\citenamefont {Dixon}(2025)}]{dixon_cosmic_2025}%
  \BibitemOpen
  \bibfield  {author} {\bibinfo {author} {\bibfnamefont {K.~H.}\ \bibnamefont {Dixon}},\ }\emph {\bibinfo {title} {A Measurement of the Cosmogenic Neutron Production in Water at the SNO+ Experiment}},\ \href {https://snoplus.phy.queensu.ca/theses/kdixon.pdf} {Ph.D. thesis},\ \bibinfo  {school} {King's College London} (\bibinfo {year} {2025})\BibitemShut {NoStop}%
\bibitem [{\citenamefont {Kudryavtsev}\ \emph {et~al.}(2003)\citenamefont {Kudryavtsev}, \citenamefont {Spooner},\ and\ \citenamefont {McMillan}}]{kudryavtsev_simulations_2003}%
  \BibitemOpen
  \bibfield  {author} {\bibinfo {author} {\bibfnamefont {V.}~\bibnamefont {Kudryavtsev}}, \bibinfo {author} {\bibfnamefont {N.}~\bibnamefont {Spooner}},\ and\ \bibinfo {author} {\bibfnamefont {J.}~\bibnamefont {McMillan}},\ }\href {https://linkinghub.elsevier.com/retrieve/pii/S0168900203009835} {\bibfield  {journal} {\bibinfo  {journal} {Nucl. Instrum. Methods Phys. Res., Sect. A}\ }\textbf {\bibinfo {volume} {505}},\ \bibinfo {pages} {688} (\bibinfo {year} {2003})}\BibitemShut {NoStop}%
\bibitem [{\citenamefont {Araújo}\ \emph {et~al.}(2005)\citenamefont {Araújo}, \citenamefont {Kudryavtsev}, \citenamefont {Spooner},\ and\ \citenamefont {Sumner}}]{araujo_muon-induced_2005}%
  \BibitemOpen
  \bibfield  {author} {\bibinfo {author} {\bibfnamefont {H.~M.}\ \bibnamefont {Araújo}}, \bibinfo {author} {\bibfnamefont {V.~A.}\ \bibnamefont {Kudryavtsev}}, \bibinfo {author} {\bibfnamefont {N.~J.~C.}\ \bibnamefont {Spooner}},\ and\ \bibinfo {author} {\bibfnamefont {T.~J.}\ \bibnamefont {Sumner}},\ }\href {https://www.sciencedirect.com/science/article/pii/S0168900205005838} {\bibfield  {journal} {\bibinfo  {journal} {Nucl. Instrum. Methods Phys. Res., Sect. A}\ }\textbf {\bibinfo {volume} {545}},\ \bibinfo {pages} {398} (\bibinfo {year} {2005})}\BibitemShut {NoStop}%
\bibitem [{sup()}]{supp_mat}%
  \BibitemOpen
  \href@noop {} {}\bibinfo {note} {See Supplemental Material \url{ http://link.aps.org/supplemental/10.1103/vs3y-sbb2} for the selected muon, neutron and random background data used in the analysis.}\BibitemShut {Stop}%
\end{thebibliography}%


\begin{thebibliography}{0}%
\makeatletter
\providecommand \@ifxundefined [1]{%
 \@ifx{#1\undefined}
}%
\providecommand \@ifnum [1]{%
 \ifnum #1\expandafter \@firstoftwo
 \else \expandafter \@secondoftwo
 \fi
}%
\providecommand \@ifx [1]{%
 \ifx #1\expandafter \@firstoftwo
 \else \expandafter \@secondoftwo
 \fi
}%
\providecommand \natexlab [1]{#1}%
\providecommand \enquote  [1]{``#1''}%
\providecommand \bibnamefont  [1]{#1}%
\providecommand \bibfnamefont [1]{#1}%
\providecommand \citenamefont [1]{#1}%
\providecommand \href@noop [0]{\@secondoftwo}%
\providecommand \href [0]{\begingroup \@sanitize@url \@href}%
\providecommand \@href[1]{\@@startlink{#1}\@@href}%
\providecommand \@@href[1]{\endgroup#1\@@endlink}%
\providecommand \@sanitize@url [0]{\catcode `\\12\catcode `\$12\catcode `\&12\catcode `\#12\catcode `\^12\catcode `\_12\catcode `\%12\relax}%
\providecommand \@@startlink[1]{}%
\providecommand \@@endlink[0]{}%
\providecommand \url  [0]{\begingroup\@sanitize@url \@url }%
\providecommand \@url [1]{\endgroup\@href {#1}{\urlprefix }}%
\providecommand \urlprefix  [0]{URL }%
\providecommand \Eprint [0]{\href }%
\providecommand \doibase [0]{https://doi.org/}%
\providecommand \selectlanguage [0]{\@gobble}%
\providecommand \bibinfo  [0]{\@secondoftwo}%
\providecommand \bibfield  [0]{\@secondoftwo}%
\providecommand \translation [1]{[#1]}%
\providecommand \BibitemOpen [0]{}%
\providecommand \bibitemStop [0]{}%
\providecommand \bibitemNoStop [0]{.\EOS\space}%
\providecommand \EOS [0]{\spacefactor3000\relax}%
\providecommand \BibitemShut  [1]{\csname bibitem#1\endcsname}%
\let\auto@bib@innerbib\@empty
%</preamble>
\end{thebibliography}%
